\documentclass[fleqn,usenatbib]{mnras}

\usepackage{newtxtext,newtxmath}

\usepackage[T1]{fontenc}

\DeclareRobustCommand{\VAN}[3]{#2}
\let\VANthebibliography\thebibliography
\def\thebibliography{\DeclareRobustCommand{\VAN}[3]{##3}\VANthebibliography}

\usepackage{graphicx}	
\usepackage{amsmath}	
\usepackage{tabularx}
\usepackage{nicefrac}   

\newcommand{\msun}{\ensuremath{\mbox{M}_{\odot}}}
\newcommand{\lsun}{\ensuremath{\mbox{L}_{\odot}}}
\newcommand{\rsun}{\ensuremath{\mbox{R}_{\odot}}}

\newcommand{\teff}{\ensuremath{T_{\rm eff}}}

\newcommand{\logg}{\ensuremath{\log{g}}}
\newcommand{\vsini}{\ensuremath{v_{\rm e}\sin{i}}}
\newcommand{\kms}{\ensuremath{\mbox{km s$^{-1}$}}}
\newcommand{\zPup}{\ensuremath{\zeta~\mbox{Pup}}}

\newcommand{\Ha}{\mbox{H\ensuremath{\alpha}}}
\newcommand{\mdot}{\ensuremath{\dot{M}}}

\newcommand\hiddenref[1]{\sbox0{\ref{#1}}}

\def\orcid#1{\kern .08em\href{https://orcid.org/#1}{\includegraphics[keepaspectratio,width=0.7em]{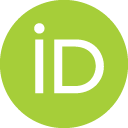}}}

\title[Polarization variations in $\zeta$~Puppis]{Rapid polarization variations in the O4 supergiant $\zeta$~Puppis}

\author[Jeremy Bailey et al.]{Jeremy Bailey$^{1}$\thanks{E-mail: j.bailey@unsw.edu.au}\orcid{0000-0002-5726-7000},
Ian D. Howarth$^{2}$\orcid{0000-0003-3476-8985},
Daniel V. Cotton$^{3,4}$\orcid{0000-0003-0340-7773},
Lucyna Kedziora-Chudczer$^{5}$\orcid{0000-0001-7212-0835},
\newauthor
Ain De Horta$^{4}$\orcid{0000-0001-9677-1499},
Sarah L. Martell$^{1,6}$\orcid{0000-0002-3430-4163},
Colin Eldridge$^{7}$,
Paul Luckas$^{8}$\orcid{0000-0001-5076-4656}
\\
\\
$^{1}$School of Physics, University of New South Wales, Sydney, NSW 2052, Australia\\
$^{2}$University College London, Gower Street, London WC1E 6BT, UK.\\
$^{3}$Monterey Institute for Research in Astronomy, 200 Eighth Street, Marina, CA, 93933, USA.\\
$^{4}$Western Sydney University, Locked Bag 1797, Penrith-South DC, NSW 2751, Australia.\\
$^{5}$University of Southern Queensland, Centre for Astrophysics, Toowoomba, QLD 4350, Australia\\
$^{6}$Centre of Excellence for All-Sky Astrophysics in Three Dimensions (ASTRO 3D), Australia\\
$^{7}$242 Randell Rd, Mardella, WA 6125, Australia\\
$^{8}$International Centre for Radio Astronomy Research, The University of Western Australia, 35 Stirling Hwy, WA 6009, Australia\\
}

\date{Accepted XXX. Received YYY; in original form ZZZ}

\pubyear{2023}

\begin{document}
\label{firstpage}
\pagerange{\pageref{firstpage}--\pageref{lastpage}}
\maketitle

\begin{abstract}
We present time-series linear-polarization observations of the bright O4 supergiant $\zeta$~Puppis. The star is found to show polarization variation on timescales of around an hour and longer. Many of the observations were obtained contemporaneously with \textit{Transiting Exoplanet Survey Satellite} ({\it TESS}) photometry. We find that the polarization varies on similar timescales to those seen in the {\it TESS} light-curve. The previously reported 1.78-day photometric periodicity is seen in both the {\it TESS} and polarization data. The amplitude ratio of photometry to polarization is $\sim$9 for the periodic component and the polarization variation is oriented along position angle $\sim$70\degr--160\degr. Higher-frequency stochastic variability is also seen in both datasets with an amplitude ratio of $\sim$19 and no preferred direction. We model the polarization expected for a rotating star with bright photospheric spots and find that models that fit the photometric variation produce too little polarization variation to explain the observations. We suggest that the variable polarization is more likely the result of scattering from the wind, with corotating interaction regions producing the periodic variation and a clumpy outflow producing the stochastic component. The H$\alpha$ emission line strength was seen to increase by 10\% in 2021 with subsequent observations showing a return to the pre-2018 level.

\end{abstract}

\begin{keywords}
polarization -- techniques: polarimetric -- stars: individual: \zPup\ -- stars: winds
\end{keywords}



\section{Introduction}
\label{sec:intro}

There is no star in the sky that is both hotter and brighter than
$\zeta$~Puppis\footnote{It is the brightest star in Puppis, with the $\alpha$--$\epsilon$ assignments being distributed among other components of the former constellation of Argo Navis.} (HD~66811, Naos);  some observational basics are listed in Table~\ref{tab:basics}.  
On the basis of
its O4\;I(n)fp classification we will refer to it as a supergiant, although it should be noted that this is a \textit{spectroscopic} designation;  in evolutionary terms, \zPup\ appears to be a product of binary evolution, and may be core-hydrogen burning \citep{howarth19}.  Additional spectral-type qualifiers indicate moderately broad lines (`(n)'),  
He\,\textsc{ii}~$\lambda$4686 and strong N\,\textsc{iii}~$\lambda$4640 emission (`f'), and unspecified peculiarities (`p').

\begin{table}
\center
\caption{Basic observed properties}
\label{tab:basics}
\begin{tabular}{lclc}
\hline
Property&\multicolumn{2}{c}{Value}&Source\\
\hline
Sp. type & O4\;I(n)fp &&1\\
$V$&2.25&&2\\
$B-V$& $-0.27\phantom{-}$&&2\\
$E(B-V)$&0.04&&3\\
\vsini&$213\pm 7\phantom{1}$&\kms&3\\
Distance&$332\pm 11$&pc&3\\
\hline
\end{tabular}
\begin{flushleft}
Sources: \\
(1) \citet{walborn10}, \citet{sota11}.
(2) \citet{johnson66}.
(3) \citet{howarth19}.
\end{flushleft}
\end{table}

As well as being bright, \zPup\ is also luminous ($\log(L/\lsun) \simeq 5.6$; \citealt{howarth19}), with a strong radiation-driven mass outflow.  These characteristics have made it a touchstone for stellar-wind studies, starting the discovery of UV P-Cygni profiles \citep{morton69} and continuing to an extensive range of observational, modelling, and theoretical studies (at the time of writing, the \textit{Simbad} database identifies 465 papers mentioning \zPup\ published from 2000 onwards).

The star is known to be variable across the observable electro\-magnetic spectrum, on a variety of timescales.
Ground-based photometry 
obtained in 1986 and 1989 by \citet{balona92} showed irregular micro\-variability (albeit with possible periods of 5.21 or 1.21~days, the former having been identified in \Ha\ spectroscopy by \citealt{moffat81}), while \cite{marchenko98} found a period of 2.56~days in \textit{Hipparcos} photometry (1989--1993). Using $\sim$1000 days of photometry from the Solar Mass Ejection Imager (SMEI) instrument on the {\it Coriolis} satellite (2003--2006), \citet{howarth14} discovered a \mbox{1.78-d} period, with an amplitude on the order of
10~mmag, varying by a factor of $\sim$2 on 10- to 100-day timescales. This period persisted in {\it BRIght Target Explorer-Constellation} ({\it BRITE-Constellation}) data  \citep[observations 2014--2015]{rami18}, and is present in \textit{Transiting Exoplanet Survey Satellite} (\textit{TESS}) observations taken in 2019 and 2021 (\citealt{burssens20}, and below).   The higher-quality space photometry also confirms the presence of stochastic variability, of comparable amplitude to the periodic (but variable) signal but with shorter timescales 
($\sim$hours; e.g., \citealt{rami18}).

We began obtaining linear photopolarimetry of \zPup\ in April~2020. The observations were taken as part of a survey of $g^\prime$-band polarizations of the 135 stars in the {\it Hipparcos} catalogue that are brighter than $3.\!\!^{\rm m}0$ and south of declination $+30^\circ$; it extends  our earlier study of 50 southern stars within 100~pc of the Sun \citep{cotton16a}. In the course of this survey several previously unknown polarization variables have been identified, including $\zeta$~Pup, for which results are presented in this paper.

\begin{figure*}
    \centering
    \includegraphics[width=\textwidth]{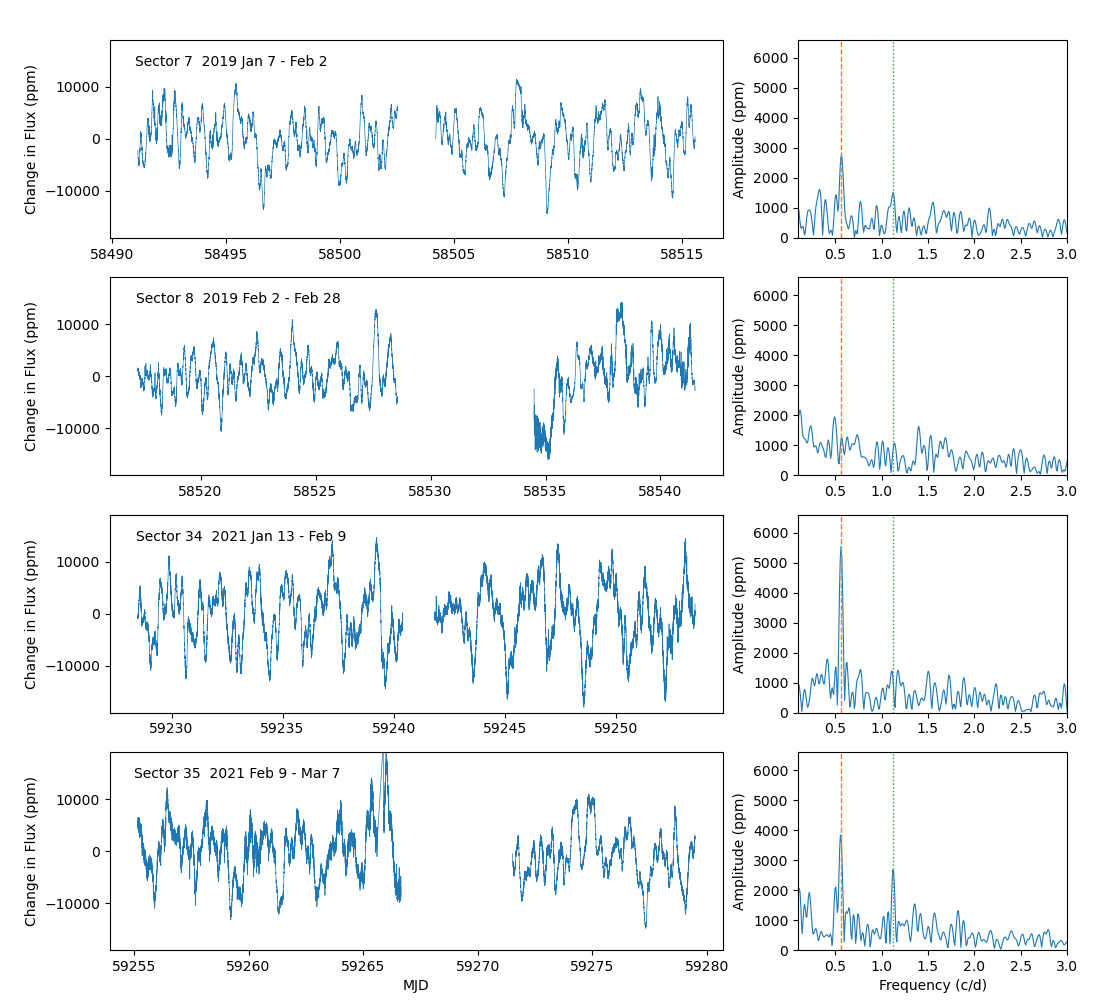}
    \caption{Light-curves (left) and periodograms (right) for $\zeta$~Pup from the four {\it TESS} sectors considered here. Light-curves are normalized PDCSAP data plotted in parts-per-million (ppm) with the mean value subtracted. The Lomb-Scargle periodograms are plotted as (semi-)amplitudes in ppm. Dashed and dotted lines are at a period of 1.78~days (0.56 c/d) and its first harmonic respectively.}
    \label{fig:tess}
\end{figure*}

\section{Observations}

\subsection{Polarization observations}

There have been few previous polarimetric observations of $\zeta$~Pup. Circular spectropolarimetry by \citet{barker81}, \citet{daviduraz14}, and \citet{hubrig16} sets a limit on any dipole magnetic-field strength
of $\lesssim$100~G. 
Linear-polarization structure through the H$\alpha$ line 
was found by \citeauthor{harries96} (\citeyear{harries96};  see also \citealt{harries00}),
who measured an $\sim$$R$-band continuum polarization of $\sim$0.04\%. 
\citet{serkowski70} reports a $V$-band polarization of $(0.09\pm0.02)$\%, while
\citet{heiles00} lists
a value of $(0.04\pm0.10)$\%, in an unspecified passband, in his agglomeration of stellar-polarization catalogues.\footnote{\citeauthor{heiles00} cites the \citet{mathewson78} compilation (CDS catalogue II/34) for this number;  they in turn record their source as ``unidentified''.    We have been unable to locate the primary source of the quoted value;  our considered speculation is that it may be an otherwise unpublished observation obtained as part of the programme described by \citet{klare72}, in which case the formal polarization uncertainty would be perhaps only $\sim$$\pm$0.02\%, at $\lambda_{\rm eff} \simeq 420$~nm.}

We obtained 255 linear-polarization observations of $\zeta$~Pup between 2020 Apr 5 and 2021 Mar 8. Most were made with the Miniature High-Precision Polarimetric Instrument \citep[Mini-HIPPI,][]{bailey17} mounted on a 23.5-cm Schmidt-Cassegrain telescope (Celestron 9.25-inch) at Pindari Observatory in Sydney. A smaller number of observations were made with the HIPPI-2 instrument \citep{bailey20a} on the 3.9-m Anglo-Australian Telescope (AAT) at Siding Spring Observatory. Both Mini-HIPPI and HIPPI-2 work in the same way, using a ferro-electric liquid crystal (FLC) modulator operating at 500~Hz and compact photomultiplier tubes as detectors.

All the Mini-HIPPI observations reported here were made using an SDSS g$^\prime$ filter. The HIPPI-2 observations were taken through three filters: a 425-nm short-pass filter (425SP), and the SDSS g$^\prime$ and r$^\prime$ filters. Transmission curves for the filters and for other instrument components are given in \citet{bailey20a}. The typical polarimetric precision achieved at g$^\prime$ on \zPup\ was $\sim$30 ppm (parts-per-million) with Mini-HIPPI and $\sim$5 ppm with HIPPI-2.

Full details of the observations and calibration methods, and tables of polarization results are given in appendix \ref{sec:polobs}.

\subsection{Space photometry}

The {\it Transiting Exoplanet Survey Satellite} \citep[{\it TESS},][]{ricker15} has observed $\zeta$~Pup in 2-minute cadence;  here we use data taken in 
sectors~7 \&~8 (2019, Jan~8 to Feb~27) and
34~\&~35 (2021, Jan~14 to Mar~6).
\textit{TESS} observes in a broad red band covering 600--1000~nm (a longer wavelength than any of the polarimetric observations).

The PDCSAP (Pre-Search Data Conditioning Simple Aperture Photometry) 
light-curves were accessed using the Python \textsc{Lightkurve}  and \textsc{Astropy} packages \citep{lightcurve18, astropy18} and are shown in Fig.~\ref{fig:tess} as normalized 
light-curves with periodograms.

\subsection{Spectroscopy}

Motivated by reports of a possible increase in 
\zPup's mass-loss rate
(\citealt{cohen20}; cf.\ Section~\ref{sec:ha}), we obtained a new spectrum of $\zeta$~Pup on 2021 Feb 2 (during the {\it TESS} sector~34 observations) using the HERMES spectrograph at the 3.9-m Anglo-Australian Telescope (AAT). HERMES has four optical bandpasses with a resolving power $R \simeq 28\,000$. For this work, we use only the CCD 3 data, which includes the \Ha\ feature. Exposure times were 0.1, 1, and 10 seconds, and the signal:noise in the combined spectrum is $\sim$90 per pixel. 
Three further spectra were obtained in Oct/Nov 2021 using a Shelyak eShel spectrograph on a PlaneWave CDK 1-m telescope at Mardella Observatory, Western Australia ($R \simeq 10^4$, fully resolving stellar spectral features). Further spectra were taken in March 2023 using the CDK 1-m telescope and a 0.35-m Ritchey-Chr\'etien telescope, also with a Shelyak eShel spectrograph, in Perth, Western Australia. 

\begin{figure}
    \centering
    \includegraphics[width=\columnwidth]{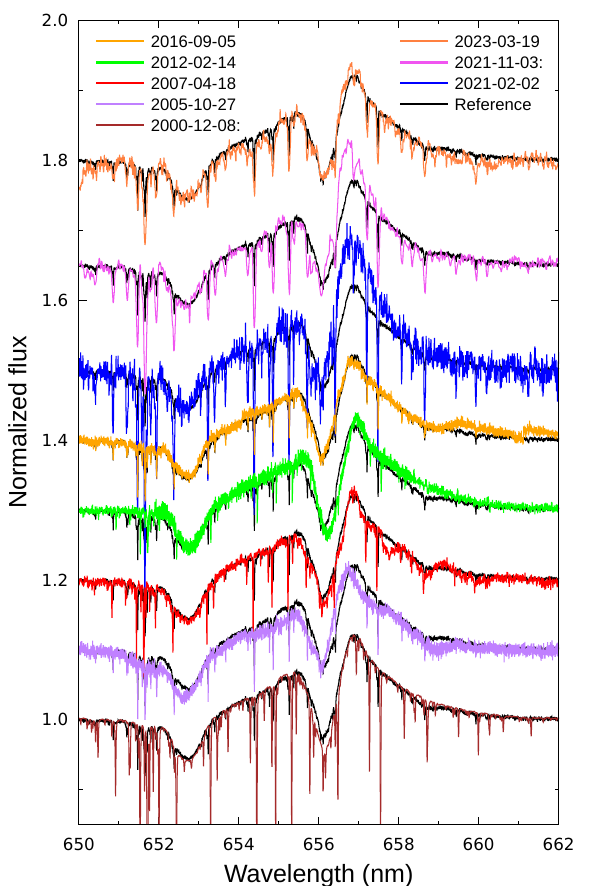}
%
    \caption{Representative H$\alpha$ spectra of $\zeta$~Pup, ordered by date, showing an enhancement in peak emission in 2021 compared with with other epochs. The spectra have been normalized to a continuum level of 1.0, and are offset in successive steps of 0.1 (0.15 for the top two spectra). Narrow absorption lines are telluric H$_2$O. The reference spectrum is the mean of observations 2000--2016; see Table \ref{tab:spec_obs} for more details.}
    \label{fig:spec}
\end{figure}

We also recovered the mean spectrum from
AAT/UCLES observations obtained in 2000 as part of an unsuccessful spectropolarimetric search for a magnetic field (Donati \& Howarth, unpublished);  and 13 further high-resolution spectra taken between 2005 and 2016, from the ESO and CFHT archives.
These echelle spectra all have resolving powers in the region of  $R \simeq 50\,000$, and signal:noise ratios in excess of 100.
A representative subset of the data, spanning the duration of the available observations, is summarized in Table~\ref{tab:spec_obs}, with the \Ha\ profiles shown in Fig.~\ref{fig:spec}.

\begin{table}
    \centering
    \caption{Selected spectroscopic observations of \zPup.}
    \begin{tabular}{l|l|l}
    \hline
    UT & Instrument & Telescope \\ \hline
    2023-03-19.48 & Shelyak eShel$^1$& CDK 1m\\
    2021-11-03$^\dagger$&Shelyak eShel& CDK 1m\\
    2021-02-02.79 & HERMES$^2$ & AAT 3.9m  \\
    2016-09-05.40 & UVES$^3$ & VLT 8.2m \\
    2012-02-14.29 & ESPaDOnS$^4$ & CFHT 3.6m \\
    2007-04-19.00 & FEROS$^5$ & ESO 2.2m   \\
    2005-10-27.27 & UVES & VLT 8.2m   \\
    2000-12-05--12$^*$ & SEMPOL/UCLES$^6$ & AAT 3.9m \\ \hline
    \end{tabular}
    \label{tab:spec_obs}
    
\flushleft
References 
1. https://www.shelyak.com,
2. \cite{sheinis16},
3. \cite{dekker00},
4. \cite{donati03},
5.~\cite{kaufer99},
6. \cite*{semel93}. \\
$\dagger$ Average of three spectra taken over 10 nights\\
*  Average of spectra taken over Dec 5--12.
\end{table}

\begin{figure*}
    \centering
    \includegraphics[width=17.5cm]{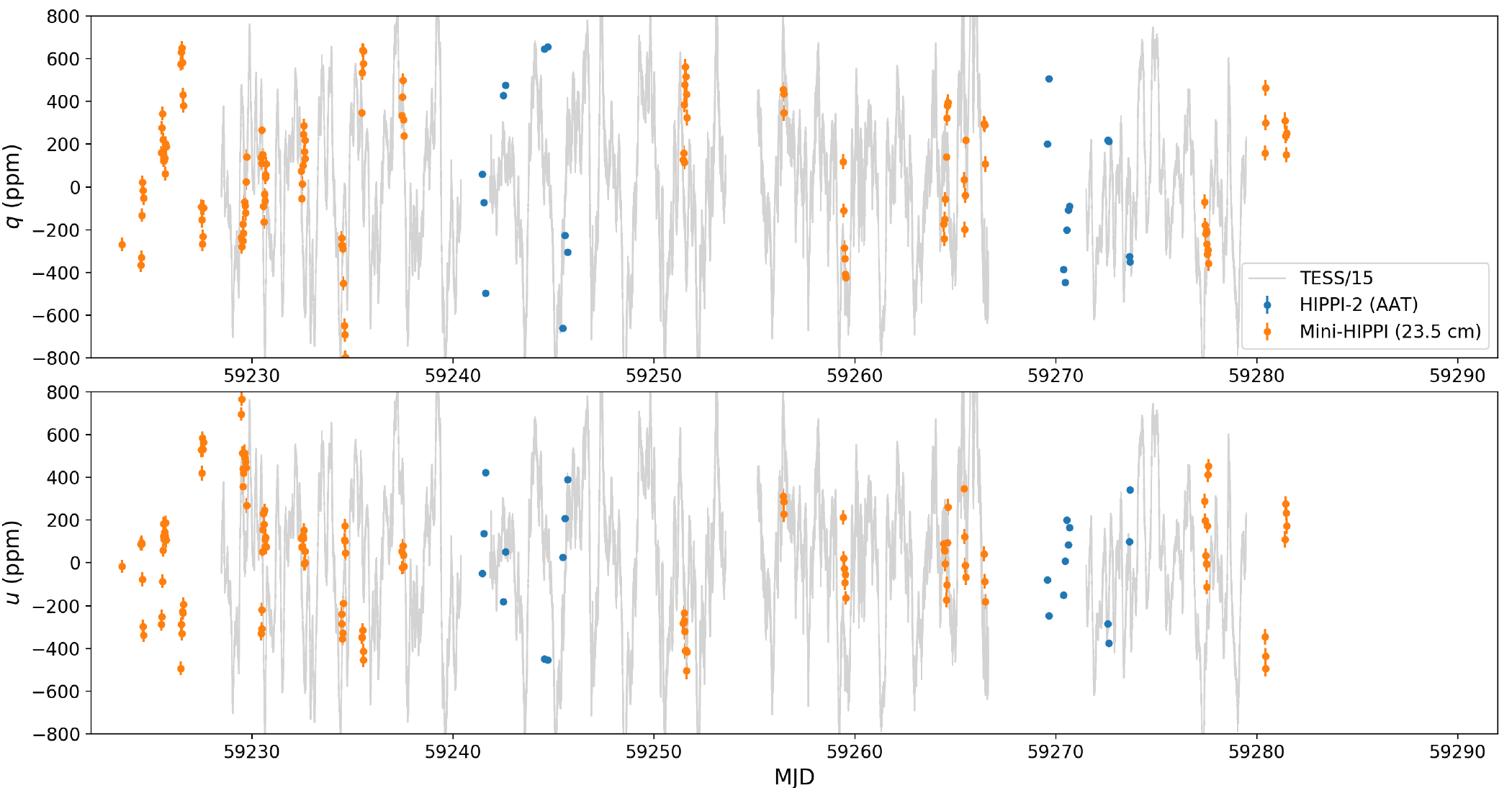}
    \caption{Polarization ($q$ and $u$) observations for times during and around the {\it TESS} sector 34 and 35 observations. The grey curves are the TESS light-curves with the mean subtracted, and divided by 15, and show that the polarization and flux vary on similar timescales.}
    \label{fig:tess_pol}
\end{figure*}

\begin{figure}
    \centering
    \includegraphics[width=\columnwidth]{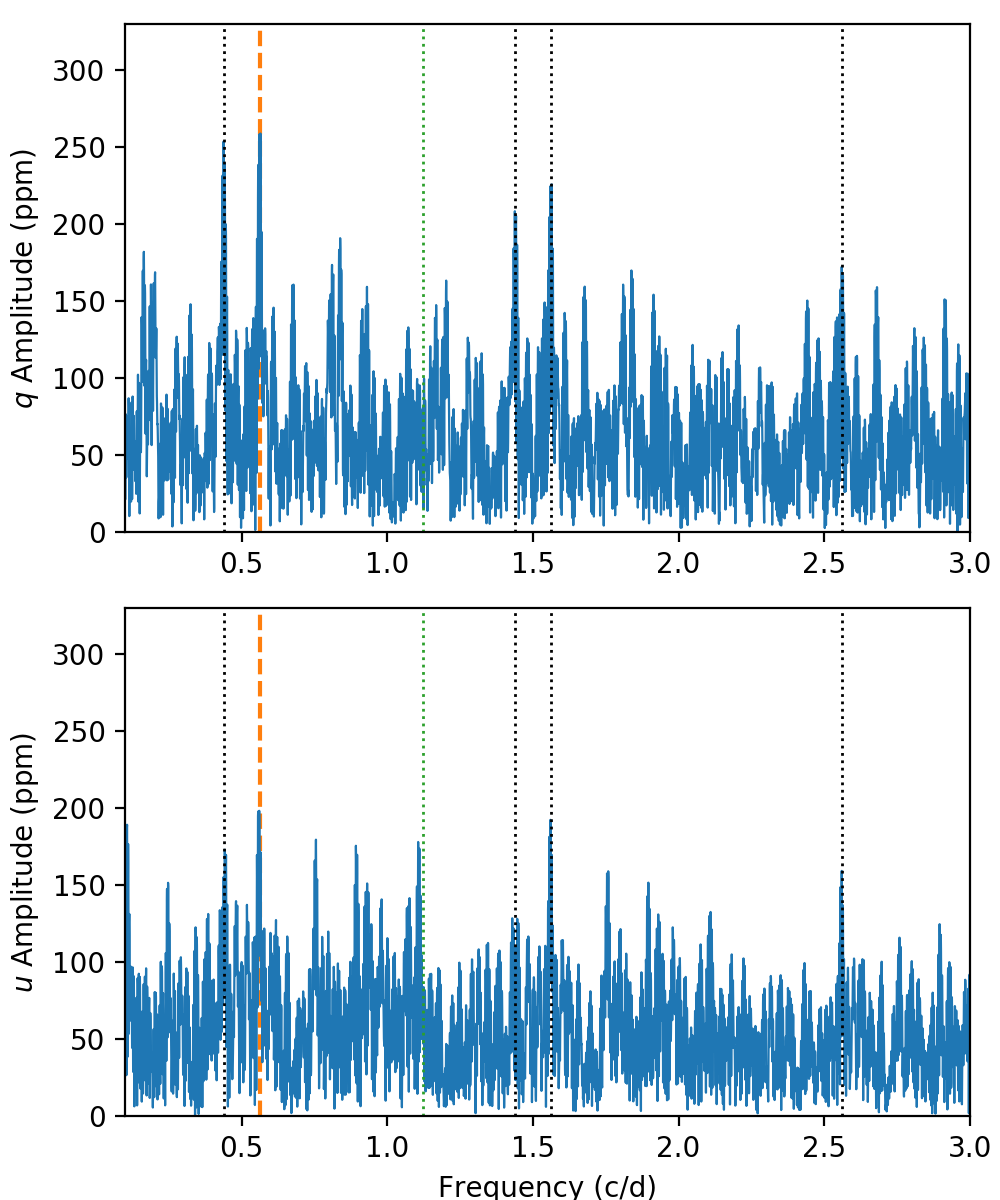}
    \caption{Periodograms of the full set of polarization ($q$ and $u$) observations (compare with those in Fig.~\ref{fig:tess}). The highest peaks in both $q$ and $u$ correspond to the 1.78 day period as marked by the dashed orange line. Other strong peaks correspond to 1 or 2 day aliases  (marked by black dotted lines)} 
    \label{fig:zpup_qu_pgram}
\end{figure}

\begin{figure}
    \centering
    \includegraphics[width=\columnwidth]{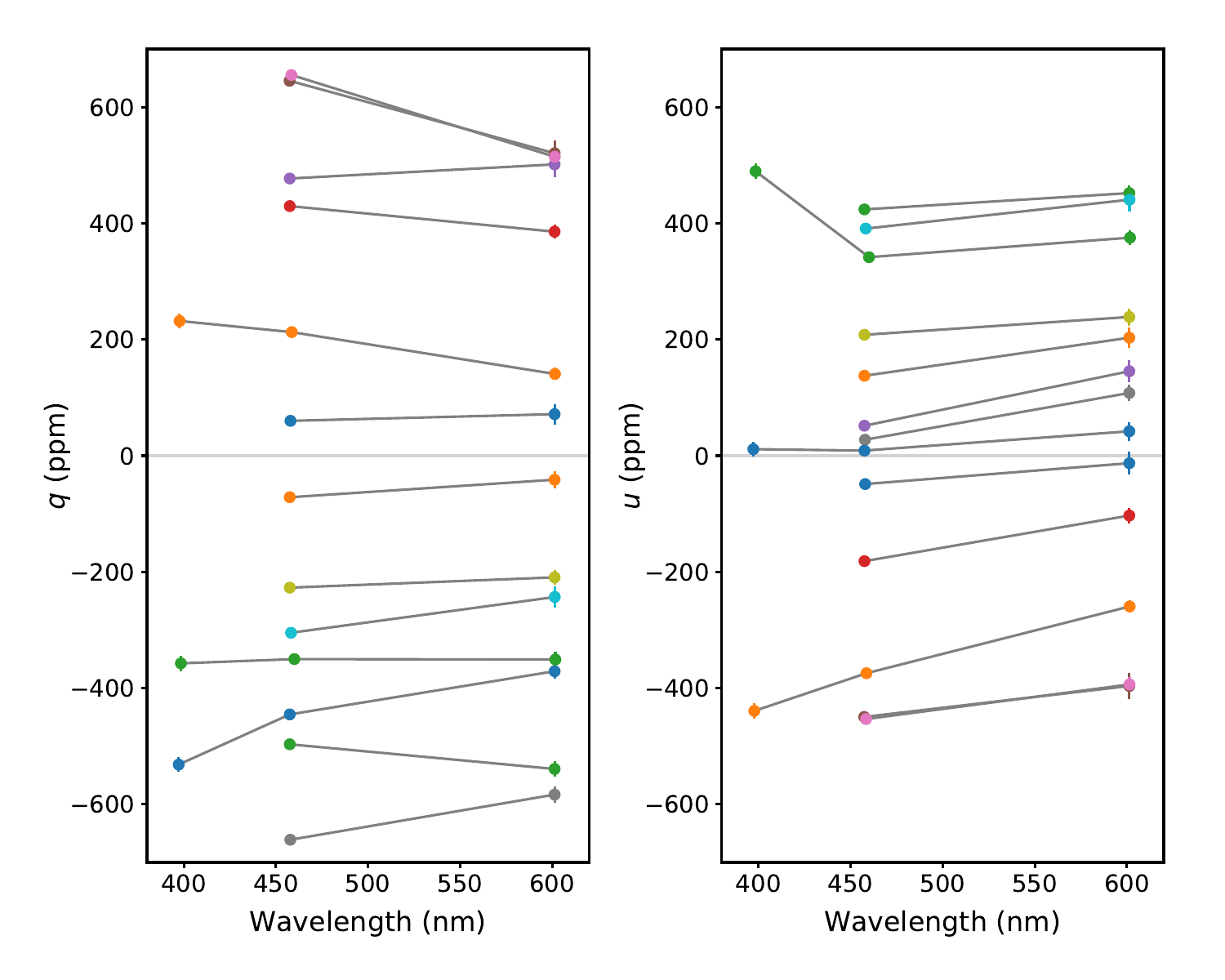}
    \caption{Polarization wavelength dependence as seen in the HIPPI-2 observations. Lines join data points obtained at closely spaced times (typically less than 30 minutes).} 
    \label{fig:zpup_wave}
\end{figure}

\begin{figure}
    \centering
    \includegraphics[width=\columnwidth]{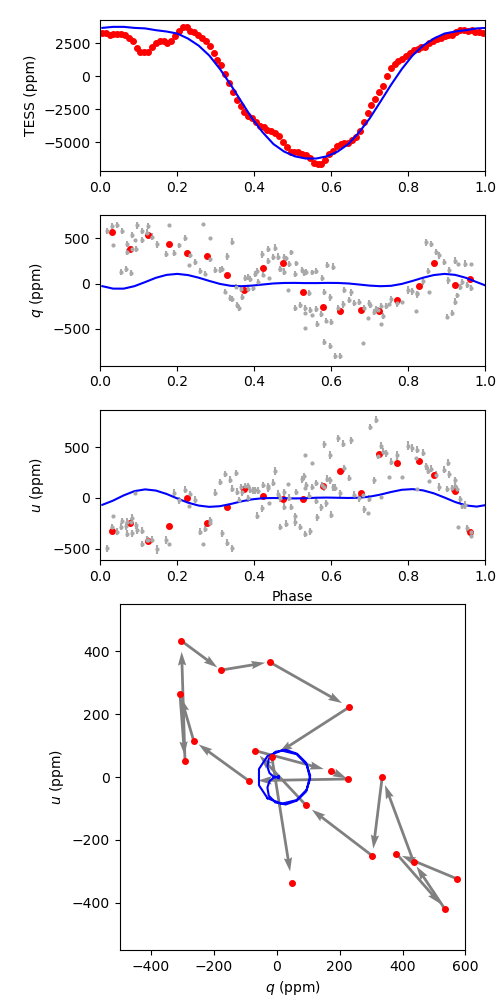}
    \caption{{\it TESS} light-curve and polarization data folded over the 1.78-day period. The top panel is the TESS light-curve for sectors~34 and~35 folded into 100 phase bins. The middle panels are the corresponding polarization data (for MJD > 59220) as individual points (grey) and averaged into 20 phase bins (red). The lower panel is the same data plotted as a QU diagram with the arrows showing the order from low to high phase values. 
    The blue lines are the predictions of a spot model as described in section~\ref{sec:rotmod}.}
    \label{fig:folded21}
\end{figure}

\begin{figure}
    \centering
    \includegraphics[width=\columnwidth]{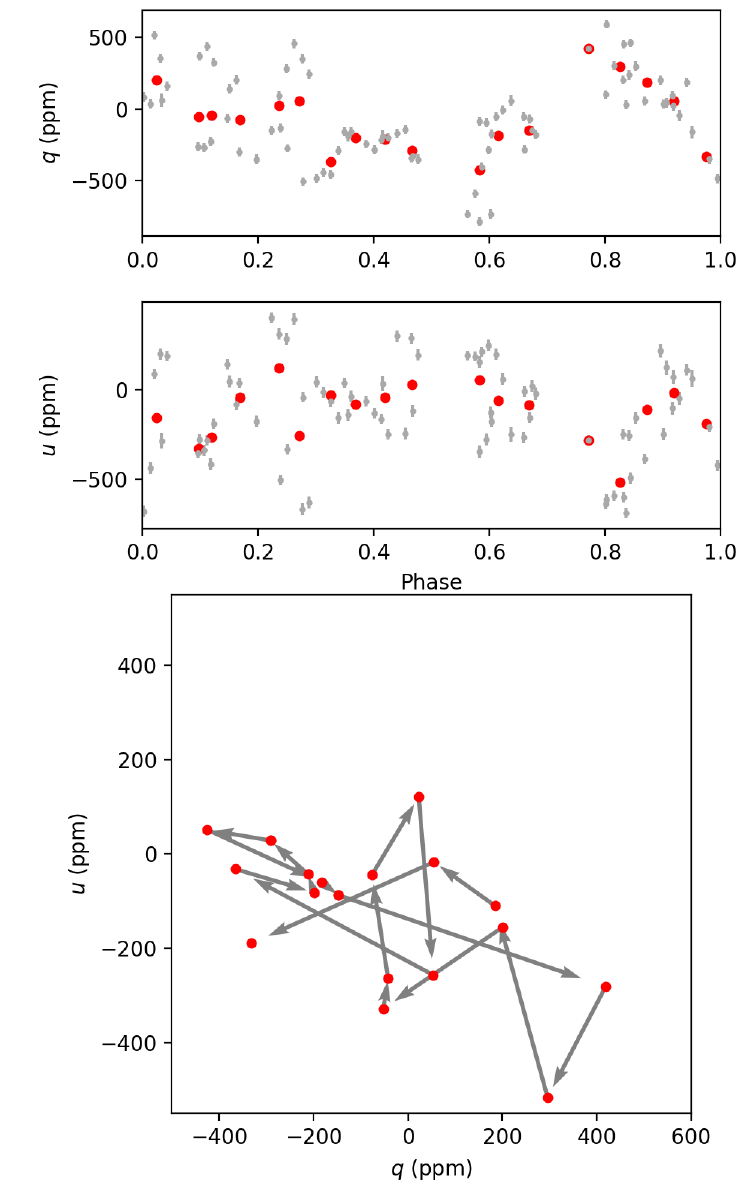}
    \caption{As Fig.~\ref{fig:folded21} but for polarization data taken in 2020 (MJD 58790 -- 59031). There are no {\it TESS} observations for these dates. Variability over the 1.78-day period is less obvious for these data.}
    \label{fig:folded20}
\end{figure}

\begin{figure}
    \centering
    \includegraphics[width=\columnwidth]{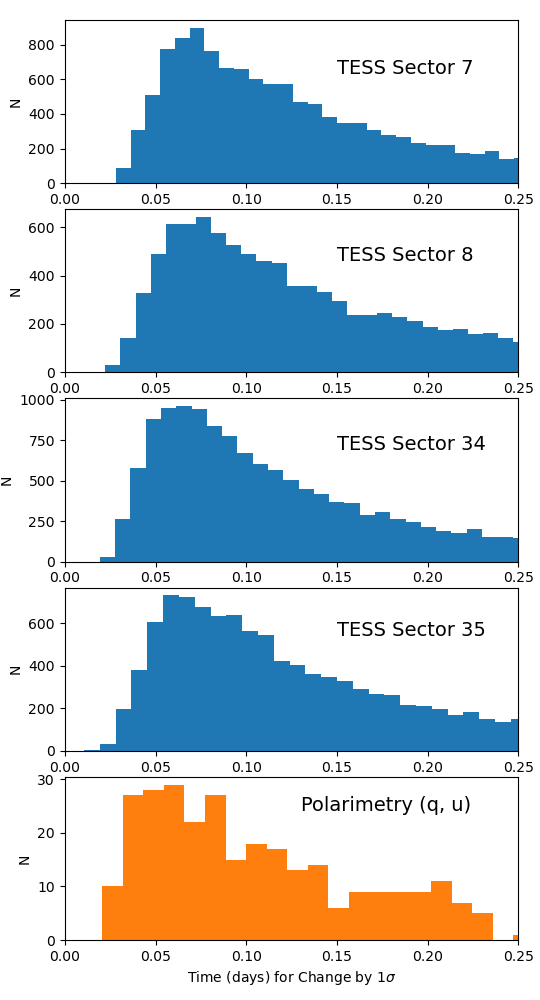}
    \caption{Timescale of stochastic variability in the four TESS sectors, and in the polarimetry. The histograms of the inverse slope of the data normalized to the standard deviations ($\sigma$) of the data points are plotted. See text for further details, and Fig. \ref{fig:tm_example} for an illustration of how the histogram data are obtained.}
    \label{fig:timescale}
\end{figure}

\begin{figure}
    \centering
    \includegraphics[width=\columnwidth]{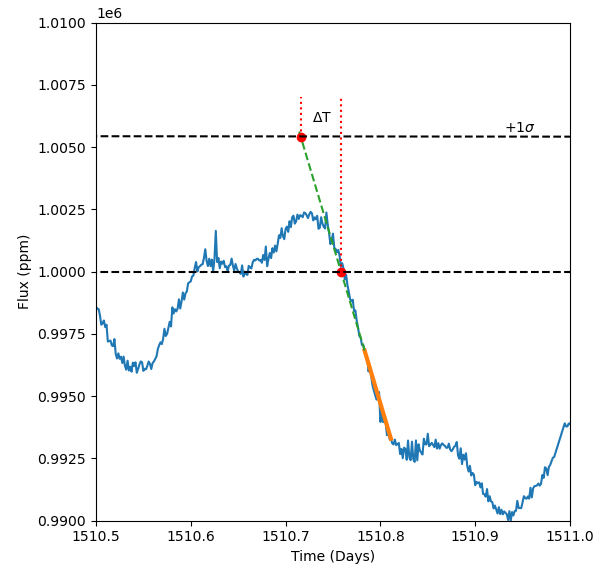}
    \caption{Illustration of the ``inverse slope'' method for obtaining the data points used in Fig \ref{fig:timescale}. The slope is measured from two points in the TESS data set 40 minutes apart (the solid orange line). The time ($\Delta$t) corresponding to a change of 1$\sigma$ is determined. The histogram of all such pairs of points is then plotted. For the polarization data, pairs of consecutive observations on the same night are used in the same way.}
    \label{fig:tm_example}
\end{figure}

\begin{figure*}
    \centering
    \includegraphics[width=18.0cm]{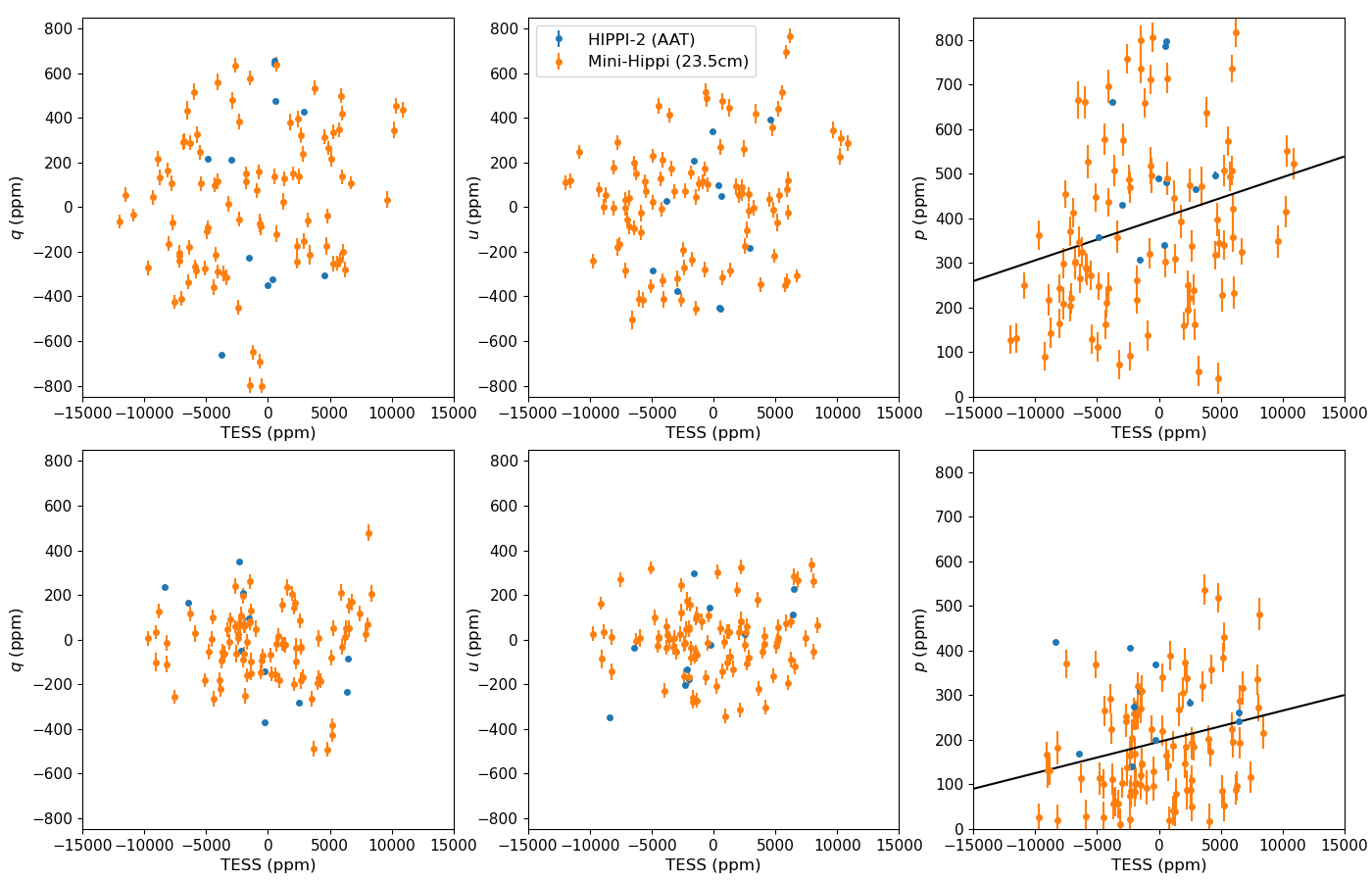}
    \caption{Correlations between polarization and photometry for polarization observations with simultaneous TESS data. Upper panels are the original observations. Lower panels have the mean light-curve and polarization curves for the periodic component (the red dots in Fig. \ref{fig:folded21}) subtracted and show only the stochastic variations.}
    \label{fig:correlation}
\end{figure*}

\begin{figure}
    \centering
    \includegraphics[width=\columnwidth]{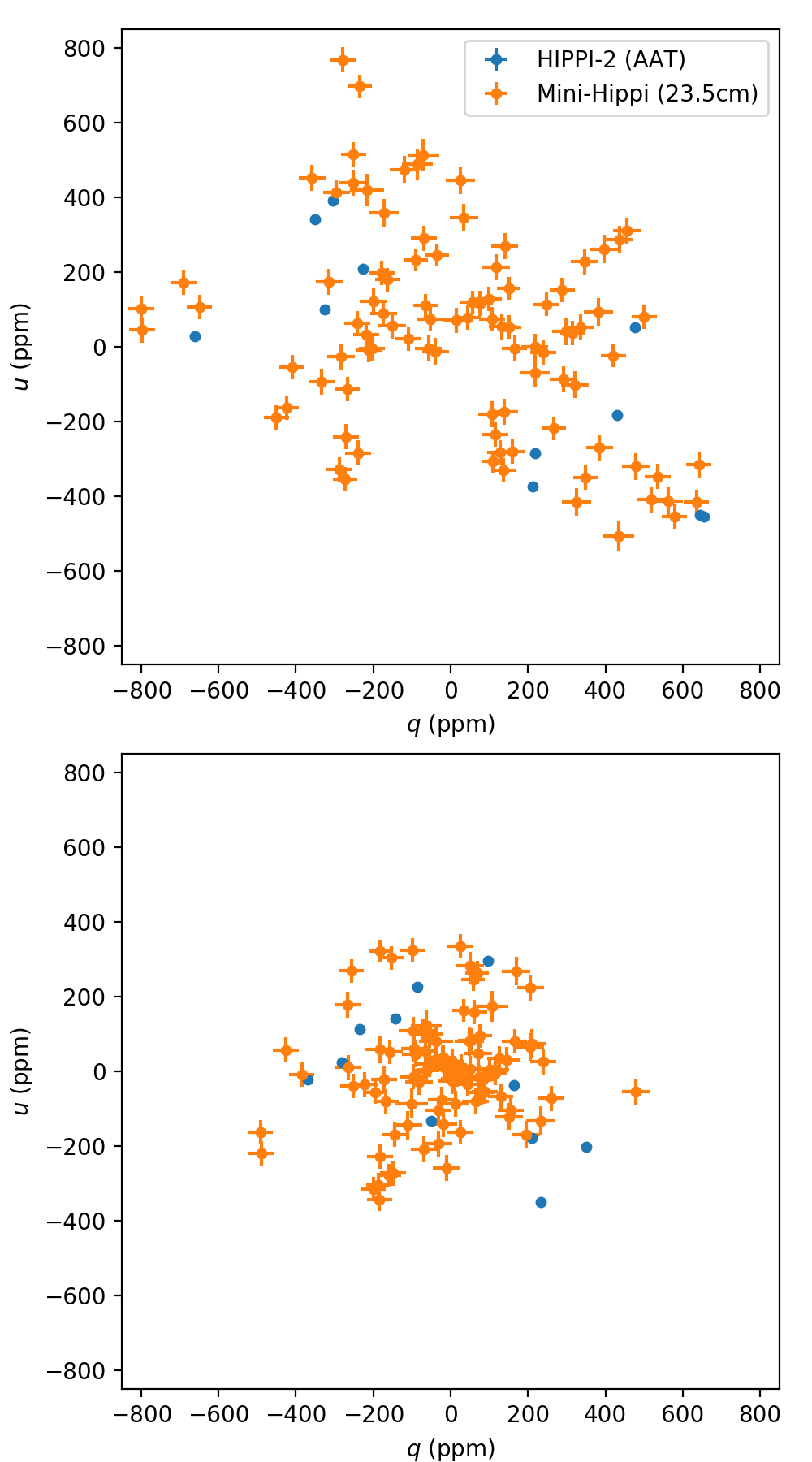}
    \caption{Correlation between normalized Stokes parameters ($q$ and $u$) for the same data sets as in Fig.~\ref{fig:correlation}. Upper panel is the original observations and shows similar structure to that seen in Fig.~\ref{fig:folded21} for the phase folded data. The lower panel has the periodic component (the red dots in Fig. \ref{fig:folded21}) subtracted to show the stochastic variations.}
    \label{fig:correlation_qu}
\end{figure}

\section{Results}

\subsection{Spectroscopy:  long-term wind changes?}
\label{sec:ha}

\citet{cohen20} reported X-ray spectroscopy of $\zeta$~Pup obtained using {\it Chandra} in 2018--2019.  They interpreted the observations as indicating a 30--40\% increase in the mass-loss rate, \mdot, compared with a similar observation from 2000. 
Because Balmer emission normally arises through recombination, the H$\alpha$ line strength is
expected to vary as density squared (in the optically thin limit, which is an adequate approximation in this case).  The suggested increase in \mdot\ would therefore give rise to an easily observable factor $\sim$2 increase in wind emission.

Modest \Ha\ variability, at the few per cent level,  is well established in \zPup;  first reported by \citet{conti76}, it has subsequently been extensively documented (e.g., \citealt{wegner78, moffat81}; \citealt{berghoefer86}; \citealt{reid96}).   As for other single, non-magnetic O-type stars (e.g., \citealt{morel04, martins15}),
the largest changes occur night-to-night, although they are observable on shorter timescales.  

We examined published \Ha\ spectra of \zPup\ spanning a half-century to review possible longer-term changes.
Sources in addition to those already cited are \citet{ebbets80, bohannan86, harries96, hillier12}, and \citet{rami18}.    As far as we can ascertain (from often small-scale plots), the peak of the profile is within 2\%\ of 1.12$\times$ continuum in all previously published spectra, as is also true for the archival spectra we have examined; the profile was `normal' as late as September 2016 (Fig.~\ref{fig:spec}).

In that context, the 2021 HERMES and eShel spectra do appear to be exceptional, with peak intensities up to $\sim$1.18$\times$ continuum;  compared to the mean of the archival spectra, the emission flux is $\sim$10\%\ greater,  measured over the velocity range $\pm$1000~\kms.   Although the increase is much less than expected on the basis of the proposed change in \mdot\ (and could arise at fixed \mdot\ from changes in the wind's velocity law, $v(r)$, or in its clumping), this does seem to offer some support for the suggestion of a change in the nature of the outflow around the time of the 2018/19 \textit{Chandra} observations, with an H$\alpha$ signature still present 1$\nicefrac{1}{2}$--2 years later.  Note that we have not been able to find any \Ha\ observations for 2018/2019.  However, the morphology appears to have returned to normal in our most recent spectrum (March 2023).

\begin{table}
\center
\caption{Results of time-series analyses of space photometry.
Semi-amplitudes are listed for the fundamental frequencies (and so may differ 
slightly from actual photometric amplitudes in the presence of significant harmonic content).}
\label{tab:tessanal}
\begin{tabular}{rccc}
\hline
\multicolumn{1}{c}{Source/Sector}&Epoch&{Period (d)}&{Semi-amp (ppm)}\\
\hline
\rule{0pt}{2.6ex} 
SMEI   &2003--6 & $1.78093\pm  0.00013$ &   $6686  \pm 492$ \\
\textit{BRITE-b} &2014/15& $1.77781 \pm 0.00076$ & $3552 \pm \phantom{0}28$\\
\textit{BRITE-r} & &$1.77778 \pm 0.00054$ & $3880 \pm 203$\\
$b+r$  &&  $1.77806\pm  0.00047$ &   $3562 \pm 148$ \\
\rule{0pt}{2.6ex} 
\textit{TESS\hfill} 7   &2019   &  $1.7658 \pm  0.0128$  &   $2808 \pm 287$ \\ 
 8   &   &  $(1.7572 \pm  0.0265\phantom{)}$  &   $\phantom{(}1266 \pm 560)$ \\ 
7+8    &  &$1.7773 \pm  0.0090$  &   $2053 \pm 385$ \\
34     &2021&  $1.7773 \pm  0.0041$  &   $5584 \pm 167$ \\
35     &  &$1.7890 \pm  0.0048$  &   $3933 \pm 400$ \\
34+35  &  &$1.7848 \pm  0.0027$  &   $4743 \pm 409$ \\
All    &2019--21&  $1.7827 \pm  0.0044$  &   $3445 \pm 399$ \\
\hline
\end{tabular}
\end{table}

\subsection{Light-curves}

The {\it TESS} light-curves and periodograms are given in Fig.~\ref{fig:tess}, with numerical results in Table~\ref{tab:tessanal}.    Previous measurements of the
\mbox{1.78-d} period, from full SMEI and
{\it BRITE-Constellation} datasets \citep{howarth14, rami18}, are included for reference.   The \textit{TESS} and SMEI results are
from a generalized Lomb-Scargle periodogram analysis (\citealt{zechmeister09}; \citealt{ferraz81}), with errors estimated from Monte-Carlo simulations using a residual-permutation algorithm.

The {\it BRITE-Constellation} results were obtained in blue and red filters (390--460~nm and 545--695~nm), with an amplitude  7$\pm$3\% greater in the blue \citep{rami18}.   The single \textit{TESS} passband has an effective wavelength $\lambda_{\rm eff} \simeq 800$~nm, and so observed amplitudes may be $\sim$10\%\ smaller than the \textit{BRITE-b} values.   SMEI also had a very broad response, with $\lambda_{\rm eff} \simeq 600$~nm for blue stars, roughly similar to the 
\textit{BRITE-r} passband.

From the periodograms it is apparent that the previously reported 1.78-day periodicity is present in the \textit{TESS} data, but is highly variable in amplitude. The signal is strongest in the sector~34 data, at an intermediate level in the sector~7 and 35 data, and is essentially absent in sector~8. The first harmonic is also apparent in sectors~7 and 35 but is variable in strength, indicating that the light-curves are variable in shape and can be non-sinusoidal. The first harmonic was also seen in the SMEI and {\it BRITE-Constellation} data reported by \citet{howarth14} and \citet{rami18}, and shows more strongly in the latter dataset than in the {\it TESS} observations.

It is clear from the light-curves that there is also substantial high-frequency variability that is not associated with the 1.78-day periodicity. This is most clearly seen in the sector-8 light-curve where the 1.78-day period is not apparent. However, similar variability is also seen in the other sectors in addition to the periodic component. There are no distinct periodicities apparent in the periodograms associated with this component. This stochastic variability was first discussed in detail by
\citet{rami18}, although \citeauthor{balona92}'s (\citeyear{balona92}) discovery of ``irregular microvariability'' at a similar level is evidently related.

\subsection{Polarization variability}

\label{sec:polvar}

The polarization variability of \zPup\ was apparent after the first few observations, made beginning in April 2020. It was soon established that substantial variation could be seen over a few hours. For example, observations on 2020 May 16 (MJD 58985) show $q$ varying from $-$687 to +54 ppm in $\sim$4 hours. 

We made intensive polarization observations during the period of {\it TESS} observations for sectors~34 and~35 in early 2021. Sequences of up to 12 observations per night were made over this period. Figure~\ref{fig:tess_pol} shows the g$^\prime$ polarization data for the normalized Stokes parameters $q$ and $u$, all given in ppm.

The polarization data are overlaid on the {\it TESS} photometery with the mean level subtracted and divided by 15. With this scaling the polarization and photometry amplitudes are similar, and it can be seen that the polarization and photometry vary on similar timescales. The plots are not intended to suggest that the photometry and polarization are correlated;  in some cases they vary in opposite directions (see discussion in Section~\ref{sec:correlation}), but it is clear that the same timescales of variability are seen in both photometry and polarimetry and that the amplitude in photometry is about 15 times that seen in each Stokes parameter.

Fig.~\ref{fig:zpup_qu_pgram} shows the periodograms of the polarization data in $q$ and $u$, using the full set of $g^\prime$-band polarization data described in Section~\ref{sec:polobs}. The highest peaks in both $q$ and $u$ correspond to the 1.78~day period. Most of the other strong peaks correspond to $\pm$1- or $\pm$2-day aliases of the 1.78~day period, as expected for the irregular spacing of these ground-based data.

Most of the polarization data were obtained only in the $g^\prime$-band, but a limited amount of data were obtained at multiple wavelengths and is shown in Fig.~\ref{fig:zpup_wave}. These show that variability in the $g^\prime$ and $r^\prime$ bands are very similar; there is perhaps a slightly increased amplitude in the blue (425SP) filter, although there are very few points.

In Fig.~\ref{fig:folded21} we show the {\it TESS} light-curve for sectors~34 and~35 and the corresponding polarization data plotted against phase. The epoch and period used for the phase determinations were, E = MJD 59230.0, P = 1.78~days. The periodicity is clearly seen in the light-curve and both Stokes parameters. The amplitudes of the binned polarization variation are 440~ppm in $q$ and 426~ppm in $u$. The amplitude in $p$, measured as half the vector distance in the $qu$ plane between the furthest two bins, is 598 ppm.
The photometric amplitude from the binned phase curve is 5184~ppm (4.8~mmag), intermediate between previously reported results (Table~\ref{tab:tessanal}). 
The ratio of photometric to polarimetric amplitude is therefore $\sim$9 for the periodic component.

Fig.~\ref{fig:folded20} shows the phase-folded polarization data taken in 2020. There is no {\it TESS} photometry at this time. There is still evidence of periodic polarization variability but the amplitude is smaller than in the 2021 data, indicating that, as for the photometry, the amplitude of 1.78-day polarization signal changes. The phasing of the polarization maximum and the form of the phase curve also look different to the 2021 data, which is again consistent with the changes 
seen in photometry.

The lower panels of Figs.~\ref{fig:folded21} and~\ref{fig:folded20} show the polarization variation over the 1.78-day period in the $qu$ plane. While there is quite a lot of scatter, the data points in Fig.~\ref{fig:folded21} (2021 data) mostly lie along a diagonal line from top left to bottom right. This corresponds to polarization position angles of $\sim$70\degr\  and $\sim$160\degr.
Figure~\ref{fig:folded20} (2020 data) shows a similar pattern but rotated a little to higher angles.

\subsection{Timescale of stochastic variability}

\label{sec:timescale}

As already noted stochastic variability is present in both the photometry and polarimetry. In Fig. \ref{fig:tess_pol} it can be seen that the variability timescales are similar by comparing the typical rise or fall time of observations on timescales of a few hours. To make this analysis more quantitative we have plotted histograms of timescales determined from the local inverse slope of the photometric or polarimetric data. These are given in Fig. \ref{fig:timescale}, with the method further illustrated in Fig. \ref{fig:tm_example}. 

For the {\it TESS} data, each point included in the histogram is derived by measuring the local slope of the photometric data from a pair of points 40 minutes apart. This is then converted to a time ($\Delta$t), which is the time corresponding to a change of 1$\sigma$ at that slope, where $\sigma$ is the standard deviation of the flux values in the full {\it TESS} datasets (as in Fig. \ref{fig:tess}). These histograms are similar for all four {\it TESS} sectors. They have a peak at about 0.06 days (1.44 hours), a minimum value of about 0.03 days (corresponding to the steepest changes seen in the light curves) and a long tail of larger values which correspond to pairs of points on the turnarounds between the rising and falling sections.

It is apparent that these histograms characterize the stochastic component of the variability and are not very sensitive to the 1.78 day periodic components, since the histograms have very similar shape in Sector 8, which has no periodic variability, and in Sector 34, which has the strongest periodic component.

The bottom panel of Fig. \ref{fig:timescale} shows the histogram derived in the same way from the polarimetry data. This is measured from pairs of consecutive polarization measurements made on the same night. The typical spacing of these data points is similar to the 40 minute value used with the TESS data. The slope was measured separately for the q and u data, and both sets were combined in the histogram. The polarimetry histogram is noisy due to the availability of far fewer pairs of observations, and the larger relative errors on the points. However, it can be seen that the shape is generally similar to that seen in the {\it TESS} photometry. The peak in the histogram appears in roughly the same place but is broader. The broadening can be understood as due to the errors on the polarimetry which can be comparable to the differences between adjacent points, and thus have a larger effect than in the case of photometry.

\begin{table}
    \caption{Standard deviations and correlation coefficients of TESS photometry and polarimetry (data points as in Fig.~\ref{fig:correlation}). }
    \centering
    \begin{tabular}{l|cc}
         &  Original & Periodic Component \\
         &  Data & Subtracted \\ \hline
  $\sigma_{TESS}$ (ppm)  &   5281\phantom{.00} & 4402\phantom{.00} \\
  $\sigma_q$ (ppm)     &   \phantom{0}334\phantom{.00} &  \phantom{0}170\phantom{.00} \\
  $\sigma_u$ (ppm)     &   \phantom{0}274\phantom{.00} &  \phantom{0}152\phantom{.00} \\
  $\sigma_p$ = $(\sigma_q^2 + \sigma_u^2)^{0.5}$ & \phantom{0}432\phantom{.00} & \phantom{0}228\phantom{.00} \\
  $\sigma_{TESS}/\sigma_p$ &   \phantom{00}12.2\phantom{0} & \phantom{00}19.3\phantom{0} \\
  corr $(q,TESS)$       &   \phantom{$-$0}0.19 &  \phantom{0}$-$0.05 \\
  corr $(u,TESS)$       &   \phantom{$-$0}0.22 &   \phantom{$-$0}0.11 \\
  corr $(p,TESS)$       &   \phantom{$-$0}0.25 &  \phantom{$-$0}0.25 \\
  corr $(q,u)$          &   \phantom{0}$-$0.42 & \phantom{$-$0}0.05 \\ \hline
    \end{tabular}
    \label{tab:correlation}
\end{table}

\subsection{Correlation analysis}
\label{sec:correlation}

In Fig.~\ref{fig:correlation} we show plots of the correlation between the {\it TESS} photometry and the polarization data ($q$, $u$, and $p = \sqrt{q^2+u^2}$) for the 104 polarization observations taken during {\it TESS} coverage. The {\it TESS} photometry has been averaged over 30-minute windows centred on the mid-point time of each polarization observation to provide the corresponding values plotted in the figures. The three top panels in Fig.~\ref{fig:correlation} show the original data that include the periodic as well as the stochastic variability. In the lower panels of Fig.~\ref{fig:correlation} the phase binned curves for the periodic components (as shown by the red dots in Fig.~\ref{fig:folded21}) have been subtracted from the data points for both photometry and polarimetry to leave only the stochastic component.

Table \ref{tab:correlation} presents the standard deviations of the data points plotted in Fig.~\ref{fig:correlation} and the Pearson correlation coefficients, $r$, corresponding to each of the six panels in the figure. Based on these standard deviations the ratio of variability amplitudes in photometry and polarization is 12.2 for the total variability and 19.3 for the stochastic component alone. In Section~\ref{sec:polvar} we found an amplitude ratio of 9 for the periodic component alone. The periodic component of the variability thus shows up more strongly in polarization. Nevertheless, it is clear from the lower panels of Fig.~\ref{fig:correlation} that the stochastic component is clearly seen in polarization. The scatter in the data points is many times larger than the statistical errors on the data points. This is shown in particular by the AAT HIPPI-2 data (blue points on the plot) for which the errors are typically $\sim$5 ppm, whereas the points scatter over $\pm$200 ppm.

The correlation coefficients are also listed in Table~\ref{tab:correlation}. For our sample size, $r$ values greater than 0.193 [0.251] are significant at the 5\% [1\%] level. For the original data, which includes the periodic component, we expect to see some correlation given that the periodic signal is present in all datasets (as shown in Fig.~\ref{fig:folded21}). The actual measured correlations between polarization and photometry are not strong ($|r| \simeq 0.19$--0.25). This can be understood as a result of the different phasing and shapes of the phase curves.

For the data sets that have the periodic component subtracted (lower panel of Fig.~\ref{fig:correlation}) we find no significant correlation between $q$, $u$, and the {\it TESS} photometry (correlation coefficients of $-0.05$ and $0.11$). However, there is a larger correlation ($r = 0.25$, significant at the 1\% level) between $p$ and photometry. The corresponding regression line is shown in the lower-right panel of Fig.~\ref{fig:correlation}, and shows that, typically, $p$ increases from $\sim$100~ppm for the faintest points to $\sim$220~ppm for the brightest, although there remains a large scatter around this line. Fig.~\ref{fig:correlation_qu}, which shows the correlation between $q$ and $u$ for the same datasets as in Fig.~\ref{fig:correlation}, helps to explain what we are seeing. The lower panel in this figure shows that there is no preferred direction for the stochastic polarization variations. The correlation coefficient is 0.05. The data seem consistent with the stochastic variability being due to a series of events each of which increases the brightness and polarization but with random orientations, leaving no correlations with $q$ and $u$.

In contrast the upper panel of Fig.~\ref{fig:correlation_qu} for the full dataset shows the clear preferred orientation from top left to bottom right, as already noted in Section~\ref{sec:polvar} and shown in Fig.~\ref{fig:folded21}. The correlation coefficient between $q$ and $u$ here is $-$0.42, the largest of any of the correlation plots.

\section{Discussion}

The new observations reported here, and in particular the detection of polarization variability associated with both the periodic and stochastic components of the variation, provides some additional constraints on the causes of the variability. Two mechanisms that have been discussed as explanations for the periodic variability are pulsation and rotational modulation arising from surface inhomogeneities (`starspots').

\subsection{Polarization modelling}

\label{sec:polmod}

\begin{figure}
    \centering
    \includegraphics[width=\columnwidth]{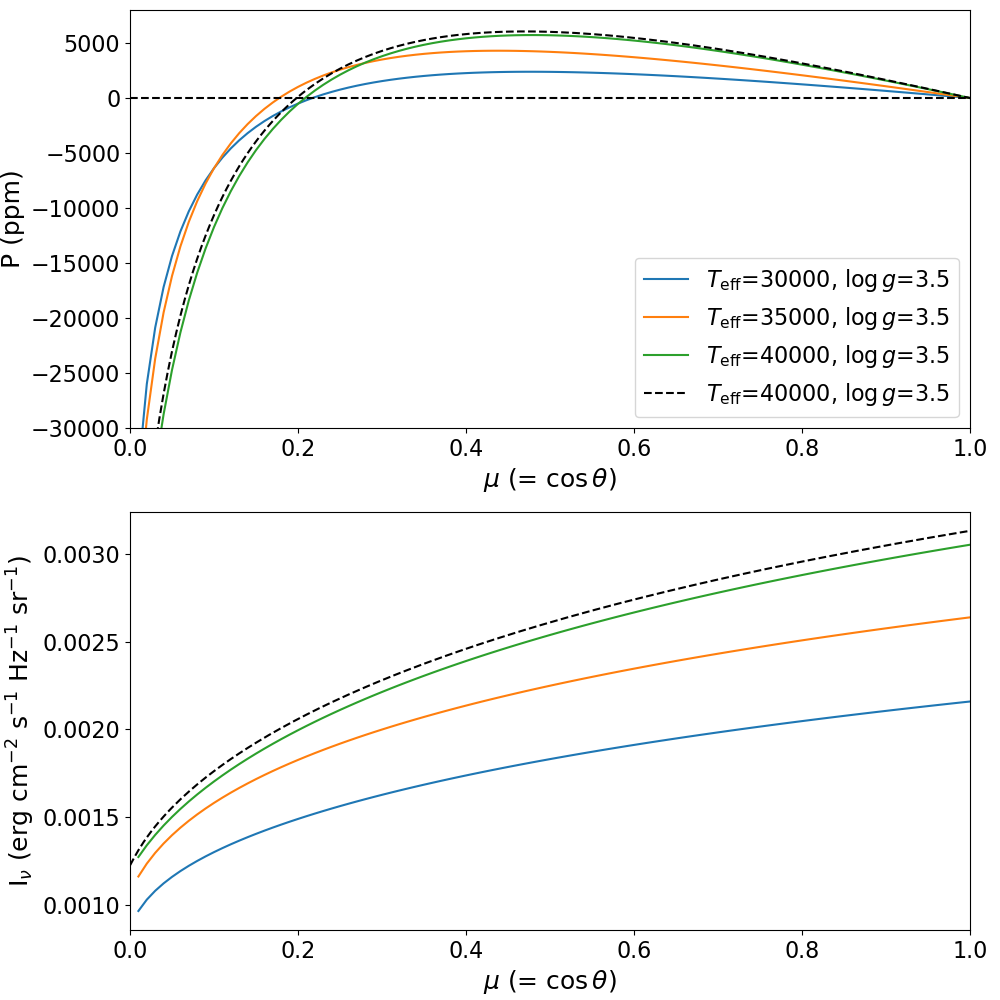}
    \caption{Predicted polarization and specific intensity at 460 nm as a function of $\mu$ for OSTAR2002 stellar-atmosphere models at $\logg = 3.5$, as described in section~\ref{sec:polmod}. Positive polarization is perpendicular to the limb of the star, negative polarization is parallel to the limb. The solid lines are calculated with \textsc{synspec}/\textsc{vlidort}. The dashed line is a polarization calculation by \citet{harrington15} for one of the same model atmospheres.}
    \label{fig:polar35}
\end{figure}

\begin{figure}
    \centering
    \includegraphics[width=\columnwidth]{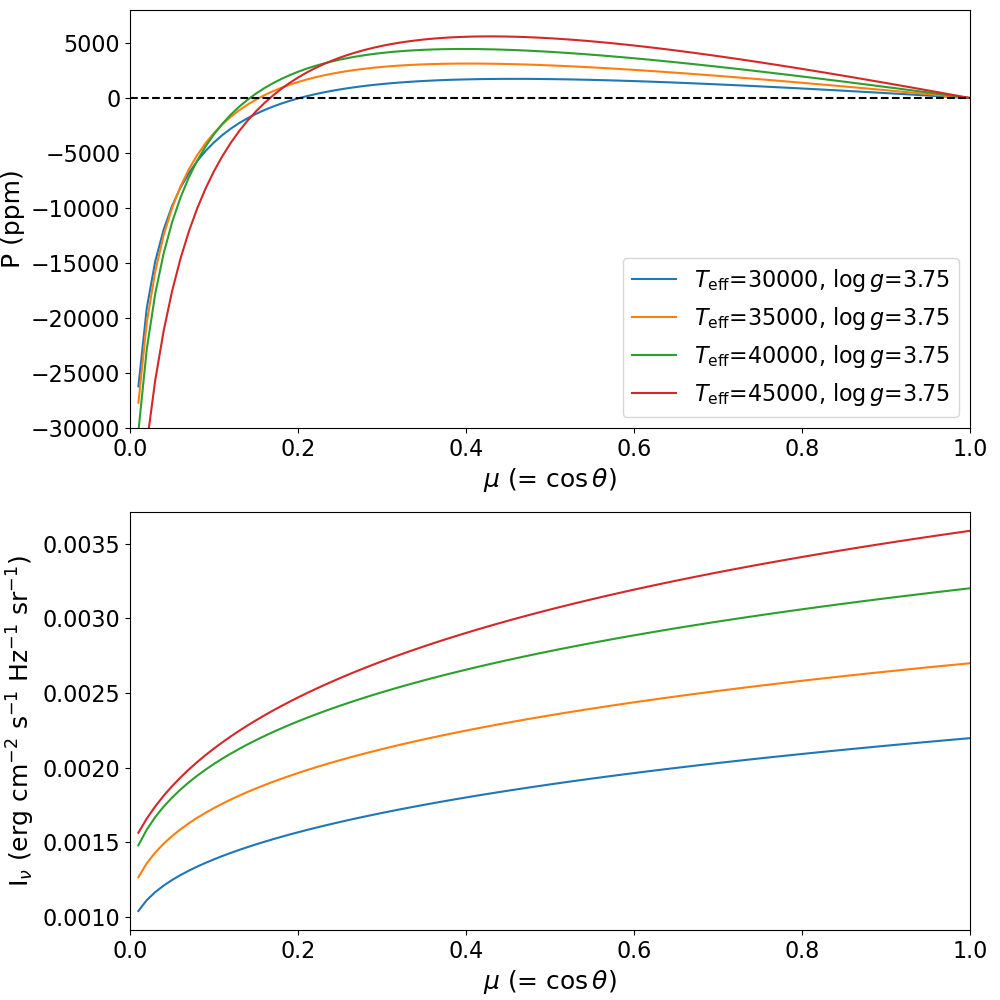}
    \caption{As Fig.~\ref{fig:polar35} for models with $\log{g}$ = 3.75.}
    \label{fig:polar375}
\end{figure}

We model the polarization produced in stellar photospheres using a version of the \textsc{synspec} spectral synthesis code \citep{hubeny85,hubeny12} modified to do a fully polarized radiative-transfer calculation, using the \textsc{vlidort} code of \citet{spurr06}. In a hot star the polarization is due to scattering from electrons. For a spherical star the radial symmetry means that the polarization will average to zero. Net polarization will arise only when there is a departure from spherical symmetry. In previous work we have observed and modelled the polarization in hot stars due to departure from spherical symmetry as a result of rotational distortion \citep{cotton17,bailey20b,lewis22,howarth23}, reflection in binary systems \citep{bailey19a,cotton20} and non-radial pulsation \citep{cotton21}.

The input required to \textsc{synspec}/\textsc{vlidort} is one or more stellar-atmosphere model structures. In past work we have generally used \textsc{atlas9} models, which can be easily calculated for the required combinations of \teff\ and $\log{g}$. The parameters of \zPup, $\teff \simeq 40$~kK, $\log{g} \simeq 3.5$, put it outside the range of \textsc{atlas9} grids \citep{castelli03,howarth11}, as the star is too close to the Eddington limit for stable models to be obtained.

Here, we instead use the OSTAR2002 grid of non-LTE models by \citet{lanz2003}, which includes models close to the Eddington limit, including the range relevant to  \zPup. The validity of using a hydrostatic model to represent hot, low-gravity stars is discussed by \citet{lanz2003}; in a real star radiation pressure results in a strong stellar wind. 
However, to test the hypothesis that \textit{surface} features (spots, or pulsation) give rise to the photometric and polarization variability these photospheric models should suffice (as argued by \citealp{rami18}). 

Examples of the polarization predicted by some of these models are given in Figs~\ref{fig:polar35} and \ref{fig:polar375}. Here the polarization and specific intensity are plotted as a function of $\mu = \cos{\theta}$ where $\theta$ is the surface-normal viewing angle. The polarization is positive (perpendicular to the limb of the star) over most of the range, but becomes large and negative (parallel to the limb) for small values of $\mu$ \citep[close to the limb of the star,][]{collins70}. Polarization for some of these model atmospheres has also been calculated using a different method by \citet{harrington15}. The dashed line in Fig.~\ref{fig:polar35} is the \citeauthor{harrington15} model for 
$\teff = 40$~kK and $\log{g} = 3.5$. Harrington included only continuum opacities in his calculations, whereas our models also include lines, which may account for the very slightly higher intensities and polarization in the \citet{harrington15} model.

From the models plotted in Fig.~\ref{fig:polar35} and \ref{fig:polar375} it is possible to see the dependence of polarization on \teff\ and gravity. Polarization increases with higher temperatures and lower gravities, a trend also seen in cooler-star models \citep{cotton17,bailey20b}. Typically polarization at 460~nm increases by factors of 2.5--3 going from 30kK to 40kK.

\subsection{Pulsation}

\label{sec:pulsation}

\begin{table}
    \centering
    \caption{Pulsation model comparison with $\beta$ Cru (see Section~\ref{sec:pulsation})}
    \begin{tabular}{l|lll|l}
    \hline
         &  \multicolumn{3}{c|}{\zPup} & $\beta$ Cru \\
       $\ell$  & $z_{l\lambda}$ & $b_{l\lambda}$ & $z_{l\lambda}$/$b_{l\lambda}$ &  $z_{l\lambda}$/$b_{l\lambda}$ \\ \hline
       1 & & \phantom{$-$}0.6815 & & \\
       2 & 0.00484 & \phantom{$-$}0.2752 & \phantom{$-$}0.0176 & \phantom{$-$}0.00406 \\
       3 & 0.01331 & \phantom{$-$}0.01842 & \phantom{$-$}0.7226 & \phantom{$-$}0.17808 \\
       4 & 0.01516 & $-$0.03862 & $-$0.3925 & $-$0.13907 \\
       5 & 0.00487 & $-$0.00465 & $-$1.0473 & $-$0.67700 \\
       \hline
    \end{tabular}
    \label{tab:pulsation}
\end{table}

Pulsation was suggested by \citet{baade86} and \citet{reid96} as a possible explanation for an $\sim$8.5 hour period seen in absorption line profiles. Subsequent spectroscopy does not show this period \citep{baade91,reid96} and it is not seen in the more-recent space photometry \citep{howarth14,rami18}. Pulsation in low-order ($\ell$ = 1,2) modes was suggested by \citet{howarth14} as the likely explanation for the 1.78-day variation. \citet{rami18} argue against a pulsation mechanism based on the non-sinusoidal phase variation, and the changes in shape of the light-curve. They point out that though some radial pulsators can show non-sinusoidal light-curves, the behaviour seen in \zPup, sometimes showing double-peaked light-curves, is incompatible with pulsation.

Our detection of large-amplitude ($\sim$400 ppm) polarization variations over the 1.78-day period provides additional constraints.  Polarization variations in hot stars can be produced by the distortion of the stellar photosphere due to non-radial pulsation. Such effects were suggested and modelled more than 40 years ago in the context of $\beta$ Cephei pulsators \citep{odell79,watson83}, but have only very recently been detected observationally \citep{cotton21}, in the bright star $\beta$~Crucis. The polarization arises from electron scattering in the stellar atmosphere, together with the departure from spherical symmetry due to the pulsations. Only non-radial modes of $\ell = 2$ or higher can result in polarization. Radial ($\ell = 0$) and dipole ($\ell = 1$) modes do not produce polarization variations \citep{watson83} and so can be ruled out as the source of the 1.78-day periodicity in \zPup.

The polarization amplitude seen in \zPup\ is about 50 times larger than that seen in $\beta$~Cru by \citet{cotton21}. However, analytic modelling such as that of \citet{watson83} predicts only the relative amplitudes in polarization and photometry. In the case of $\beta$~Cru the polarization amplitude (in g$^\prime$) was $35\times$ smaller than the photometric amplitude seen by {\it TESS}. The corresponding ratio for \zPup\ is $9\times$ as described in Section~\ref{sec:polvar}. 
For a given mode, this ratio is determined by the ratio of the quantities $z_{l\lambda}$ and $b_{l\lambda}$ as defined by \citet{watson83}. These quantities can be derived from a stellar-atmosphere model and are integrals over $\mu$ (the cosine of the local zenith angle) involving the emergent intensity and polarization.

Table~\ref{tab:pulsation} shows these quantities and their ratio for \zPup\ compared with the ratio calculated in the same way for $\beta$~Cru by \citet{cotton21}. The stellar-atmosphere model used for \zPup\ was taken from the OSTAR2002 grid \citep{lanz2003} for $\teff = 40$~kK, $\log{g} = 3.5$. The calculations were performed for wavelengths appropriate to our observations ($z_{l\lambda}$ at 460~nm for g$^\prime$ polarimetry; $b_{l\lambda}$ at 800~nm for TESS photometry).

The values for the ratio $z_{l\lambda}$/$b_{l\lambda}$ are 
greater by factors of 4.3 at 
$\ell = 2$ and 4.1 at $\ell = 3$ for \zPup\ compared to $\beta$~Cru. For equivalent modes and inclinations, this means the amplitude ratio in polarization relative to photometry will be greater by the same amounts \citep{watson83,cotton21}. The observed amplitude ratio of 9 for \zPup\ is therefore plausible for non-radial pulsation in  a similar mode to that observed in $\beta$~Cru.

However, we nevertheless consider pulsation to be unlikely 
as the correct explanation for the 1.78-day periodicity. It would
require a single mode with strong polarization to be the only period seen. In $\beta$~Cru, 11 frequencies were detected, with only two being seen in polarization and two others having much higher photometric amplitudes than the modes seen in polarization \citep{cotton21}. Additionally, the non-sinusoidal nature and changes in shape of the phase curve remain strong arguments against a pulsation origin.

\begin{figure}
    \centering
    \includegraphics[width=\columnwidth]{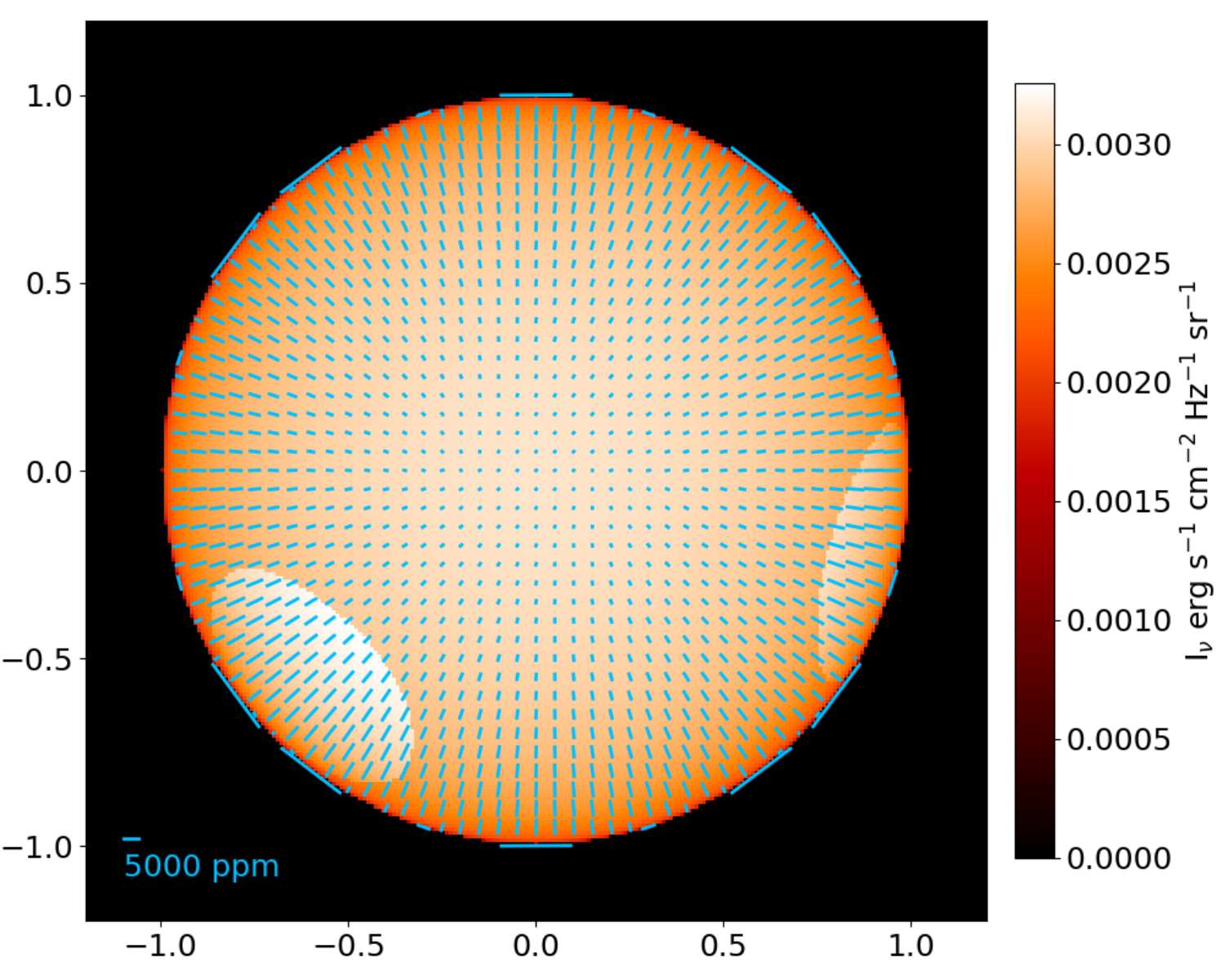}
    \caption{Example of polarization spot modelling. The distribution of specific intensity and overlaid polarization vectors are shown for the two-spot model described in section~\ref{sec:rotmod}. Without the spots the radial pattern of polarization would integrate to zero over the whole disk. The spots break this symmetry and lead to a net polarization that changes as the star rotates.}
    \label{fig:pspot}
\end{figure}

\subsection{Rotational modulation by photospheric spots}

\label{sec:rotmod}

Rotational modulation has been suggested as the cause of (different) periodicities seen in \zPup\ in spectroscopy \citep{moffat81} and photometry \citep{marchenko98}. In their analysis of the {\it BRITE-Constellation} photometry \citet{rami18} argued for the 1.78-day periodicity being due to rotational modulation. They presented models involving two bright photospheric spots that could reproduce the observed double-peaked light-curve (although, as pointed out by \citet{howarth19}, \textit{any}  low-amplitude periodic photometric signal can be reproduced by a spot model).

We here model the polarization produced by a rotating star with photospheric spots. We use a spherical model for the star, as did \citet{rami18}, ignoring the rotational flattening, which we would expect to introduce a fixed polarization offset, but not to change substantially the phase dependence (see Section~\ref{sec:rotpol}). We used OSTAR2002 model atmospheres \citep{lanz2003}, with $\teff = 40$~kK, $\log{g} = 3.75$ for the star (corresponding to the green line in Fig. \ref{fig:polar375}), and $\teff = 45$~kK, $\log{g} = 3.75$ for the spot.

To determine the integrated polarization we use a similar approach to that used by \citet{bailey20b}, overlaying a rectangular grid of pixels over the observed view of the star, with spacing 0.01 of the star radius, and calculating the specific intensity and polarization for each pixel using \textsc{synspec}/\textsc{vlidort} as described in section~\ref{sec:polmod}. This produces a map of the intensity and polarization distribution over the star such as that shown in Fig.~\ref{fig:pspot}. Summing the data over all pixels gives the integrated intensity and polarization; repeating the analysis for different rotation angles of the star enables the phase curves to be determined. The integrated intensity was calculated at 800~nm to match the TESS photometry, while the integrated polarization was calculated at 460~nm for the g$^\prime$ band polarimetry.

We used an inclination for the star's rotation axis of 33$^\circ$. This is the required value if the 1.78-day period is the rotation period, and is constrained by the measured distance, observed flux and \vsini\ as described by  \citet{howarth19}. 
The adjustable spot parameters are the number, size, and location (for fixed temperature). We tried models with single spots, and with two identical spots spaced in longitude. We were not able to reproduce the shape of the \textit{TESS} light-curve with a one-spot model. A single spot near the equator produces too small a width for the bright section of the light-curve. The width can be increased by moving the spot to higher latitude, but then the slopes of the rising and falling branches are not well matched.

A model with two spots near the equator, similar to that used by \citet{rami18}, gives a better match to the light-curve. Our {\it TESS} light-curve does not show the clear double peak seen in the {\it BRITE-Constellation} data, so the two spots need to be moved closer together in longitude. 
A good fit to the light-curve is obtained using 
a model with two circular spots at latitude 5\degr, spaced by 110\degr\ in longitude, with radii of 20.6\degr\ measured from the centre of the star. This model is shown by the blue lines in Fig.~\ref{fig:folded21}.

While this model reproduces the {\it TESS} light-curve quite well, the resulting polarization variations do not fit the observations at all. The amplitude of the $q$ and $u$ variations falls far short of that observed, and the form of the curves is quite different. Although the light-curve is single peaked, the modelled polarization curves have multiple peaks, and more structure than is seen in the observations. It should be noted that the position angle of the star's rotation axis is unknown, so the polarization model can be rotated arbitrarily in the QU plane. However, it is clear that no such rotation improves the fit to the observations. This is most obvious from the QU plot (Fig.~\ref{fig:folded21}, bottom panel), where the circular pattern produced in the model is quite unlike the extended distribution of the data points along the plot diagonal.

While our model is not unique and there are likely to be other spot configurations that can fit the light curve, there is no reason to expect such changes to result in significantly larger polarizations.

The polarization variability therefore does not seem to be consistent with an origin in photospheric spots. If the spot model is the correct interpretation for the periodic photometric variation, then the polarization variation must be produced in some other way. 

\subsection{Polarization due to corotating interaction regions}

The 1.78-day period is also seen in He\,\textsc{ii} $\lambda$4686 emission \citep{rami18} and in X-ray emission \citep{nichols21} both of which indicate that the periodicity is not confined to the photosphere, but extends at least some way into the wind. The polarization could therefore arise from scattering in material just above the photosphere, that is still corotating with the star. 

\citet*{ignace15} and \citet{carlos-leblanc19} have presented models of polarization due to corotating interaction regions (CIRs) in the wind from a hot star. CIRs are known to occur in the solar wind \citep{rouillard08} and have been invoked to explain variability in P-Cygni profiles in hot-star winds  \citep{cranmer96, morel97}. \citeauthor{cranmer96} use a simple `bright spot' model to induce azimuthal wind structure, forming spiral-like density enhancements in the wind. Scattering from the gas in these non-spherically-symmetric structures produces phase-dependent polarization. 

As a mechanism for the polarization variability, these models have the advantage that it is easier to explain the amplitude of the polarization variations, as the polarization of light scattered from electrons in optically thin gas can be very high. The mechanism also naturally results in variations repeating with the rotation period of the star. 

The models for the polarization due to a single CIR do not provide a good match to the behaviour of \zPup\ in the QU plane. However, more flexibility can be obtained by including two or more CIRs \citep{ignace15}. Using two CIRs in the wind, \citet{st-louis18} were able to fit a range of different observed polarization curves for the Wolf-Rayet star WR6 that could not be modelled with a single CIR. If the CIRs are generated by the photospheric spots used to explain the light curve, then two CIRs are to be expected for \zPup. Alternatively, if both the photometric and polarimetric periodicities are due to CIRs, then the sometimes double peaked light curve again suggests the presence of two CIRs. 

The B1Iab supergiant $\rho$ Leo (HD 91316) shows variability with a period of 26.8 days \citep{aerts18}, interpreted as ``rotational modulation by a dynamic aspherical wind''. This may be another example of the same mechanism as in \zPup, albeit with a much longer rotation period.

\subsection{The stochastic variability}

\label{sec:stochastic}

The analysis in Sections \ref{sec:polvar} and \ref{sec:correlation} shows that as well as the periodic component of variability, \zPup\ shows stochastic variability of polarization on a similar timescale and weakly correlated with the photometric variability. We note that similar variability is present in other hot stars with winds. Polarimetry of Wolf-Rayet (WR) stars \citep{St-Louis87,drissen87,robert89} has shown many of them to vary in a stochastic way with amplitudes similar to or larger than that seen in \zPup. The range of variability (measured as $\sigma_p$) is from 160 ppm (described as essentially instrumental) up to 1550 ppm in WR40 \citep{moffat91,robert89}. The equivalent value for \zPup\ is 228 ppm (Table \ref{tab:correlation}). \citet{robert89} found an anticorrelation between the polarization amplitude of these variations and the terminal wind velocity $v_\infty$. The amplitude of stochastic photometric variability in WR stars is also correlated with wind terminal velocity \citep{lenoir_craig22}. The typical ratio of photometric to polarimetric amplitude for WR stars is $\sim$20 \citep{robert92}, very similar to what we see for the stochastic component of \zPup.
Such observations suggest that the stochastic variations originate in the clumpy winds of these objects. 

Models of the polarization variability produced by a wind with optically thin clumps have been given by \citet{richardson96,li00,davies07} and \citet{li09}. The timescale of expected variability is determined in part by the wind flow time given by $R_\star/v_\infty$ where $R_\star$ is the stellar radius and $v_\infty$ is the wind terminal velocity. The time that a single clump is sufficiently close to the star to contribute to the variable polarization will be a few times the wind flow time with the factor depending on the velocity law for the wind \citep{davies07}.

For \zPup\ the wind terminal velocity is $v_\infty = 2250$ km s$^{-1}$ \citep{puls96} and the stellar radius $R_\star \sim 13\rsun$ \citep{howarth19}. The resulting wind flow time is 1.1 hours. The variability time scales we see in photometry and polarimetry ($\sim$1.4 hours, see Section \ref{sec:timescale}) are consistent with this.

These models also make predictions about the amplitude of polarization variability. The variability is expected to be proportional to the mass loss rate ($\dot{M}$) per wind flow time. This determines the amount of scattering gas close to the star. The amplitude also depends on the clump rate ($N$) the number of clumps ejected from the star per wind flow time. A large $N$ reduces the polarization variability due to statistical averaging of the random effects from many clumps. We can use the models in \cite{li00} and scale for the mass loss rate of \zPup\  \citep[$\dot{M} = 3.5 \times 10^{-6} \msun yr^{-1}$,][]{cohen10} and the wind parameters given above, to find that the $\sigma_p$ of 228 ppm would be obtained for $N \sim$ 20. However, modelling by \citet{davies07} produces polarizations about 5 times larger, and requires $N$ > 400 (the largest value modelled) to match the same observed variability. 

The models have difficulty matching the ratio of photometric to polarimetric amplitude seen in WR stars \citep{richardson96} which is observed to be $\sim$20 as also seen in \zPup. Optically thin models for the clumps produce lower ratios than this. \cite{richardson96} suggest that the clumps may be optically thick which will reduce the polarization levels due to multiple scattering, and enhance the photometric variability due to contributions from emission (in addition to scattering) from the clumps. However, no detailed modelling of winds with optically thick clumps has been performed.

One of the most variable WR stars is WR40 (HD 96548). \citet{rami19} have shown that the stochastic photometric variations of this star can be explained by a clumpy wind with the clumps scattering light from the star. This star has also been studied with simultaneous photometry and polarimetry \citep{ignace23}. The correlations between $q$, $u$ and photometry for WR 40 show some similarities to those we find for the stochastic component in \zPup\ as discussed in Section~\ref{sec:correlation}. The $q$, $u$ plots in both cases show no preferred angle. WR 40 shows no correlation between polarization and photometry. We find a small positive correlation for \zPup. The ratio of polarization to photometric amplitude is similar in both stars.

\citet{ignace23} explain the observations of WR 40 in terms of a clumpy wind model in which clumps are ejected from the star in random directions. The lack of correlation between photometry and polarization arises from the different ways in which brightness and polarization vary as the clump moves away from the star. The $q$, $u$ distribution of the stochastic variation, seen in Fig. \ref{fig:correlation_qu}, and lack of any significant $q$, $u$ correlation, indicates that the clumps are ejected with random directions, as was also the case in WR 40.

Hot supergiants have not been as well studied for polarization variability as WR stars. $\lambda$ Cephei (HD 210839), a supergiant of spectral type O6.5I(n)fp \citep{sota11}, is an example of a star that shows stochastic variation similar to \zPup\ in its {\it TESS} light-curves, and has polarization variations \citep{hayes78} of similar amplitude to those in \zPup. \citet{krticka21} use  the {\it TESS} observations of this star as an example of how stochastic variability can be generated by wind instability. Polarization variations on short timescales have also been observed in a number of OB supergiants \citep{hayes84,hayes86,lupie87}. The latter authors describe ``random polarimetric fluctuations'' in seven out of 10 objects studied which they attribute to ``electron scattering off blobs embedded in the stellar wind''.

Stochastic photometric variability\footnote{often referred to as {\it stochastic low-frequency variability}}, similar to that observed in \zPup, has been found from recent space photometry, to be common in the light curves of many OB supergiants \citep[e.g.][]{pedersen19,bowman19,burssens20}. The cause of this variability is the subject of debate, with possible stellar mechanisms being internal gravity waves \citep{bowman19} or subsurface convection \citep{cantiello21}. However, such processes, where the light variation originates at the stellar photosphere, are not expected to result in significant polarization.
For \zPup\ and other cases where the stochastic variability is seen in both photometry and polarimetry the clumpy wind model seems more plausible as the direct cause of the variability. However, this does not rule out other processes in the star, such as those just described, contributing indirectly by driving the formation of clumps in the wind.

\begin{figure*}
    \centering
    \includegraphics[trim={3.5cm 1cm 3.5cm 0.5cm}, width=\textwidth]{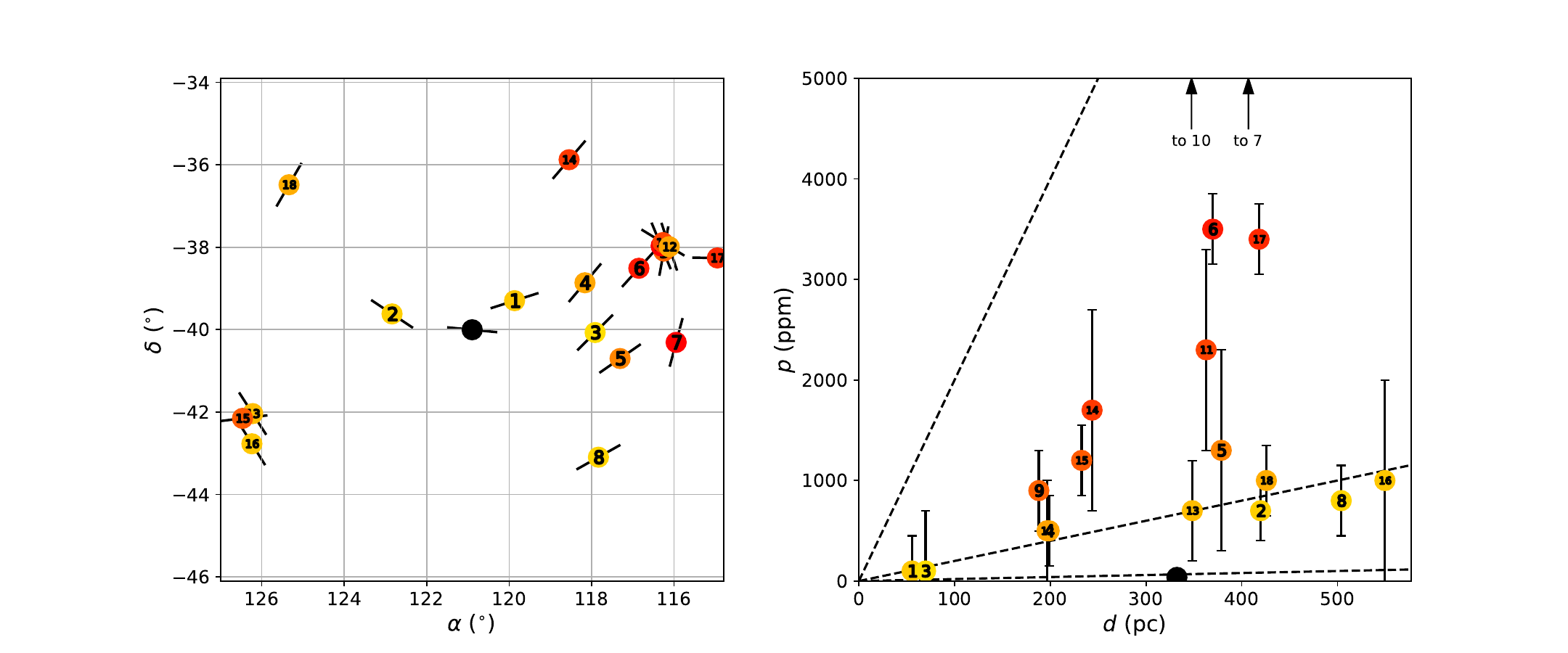}
    \caption{Observations of stars nearby to \zPup\ from \citet{heiles00} agglomerated polarization catalogue. The left hand panel shows the polarization position angle (North over East) for each star as a headless vector on a sky map. The right hand panel shows polarization as a function of distance, with dashed guidelines drawn at $p/d=$0.2, 2 and 20 ppm/pc. Plotted are all observations of stars fall within 600~pc of the Sun and 5 degrees separation of \zPup\ -- which is shown in black. The stars are colour coded according to $p/d$ and numbered in order of angular separation from \zPup\: 1: HD~65925, 2: HD~68553, 3: HD~64316, 4: HD~64503, 5: HD~63868, 6: HD~63465, 7: HD~62753, 8: HD~64287, 9: HD~62974, 10: HD~63032, 11: HD~62991, 12: HD~62876, 13: HD~71286, 14: HD~64802, 15: HD~71459, 16: HD~71302, 17: HD~61899, 18: HD~70556. The distance information for each star has been updated from the catalogue value using SIMBAD -- resulting in the use of either \citet{gaia22} or \citet{vanleeuwen2007} parallaxes. Not shown on the right hand plot are stars 10 and 7, which have polarizations of 12400 $\pm$ 350 ppm and 7200 $\pm$ 1000 ppm respectively. These two stars are known, by us, to have large intrinsic variable polarizations; the latter because it is a Be star, the former is the subject of an ongoing program of study.  A floor in polarization with distance of around 2 ppm/pc, is indicated. However, polarization increases more steeply -- approaching the 20 ppm/pc found by \citet{behr59} -- beginning from a distance of around 200~pc in a region centred around [116$^\circ$, $-$38$^\circ$]. The mean polarization of \zPup\ is well below that which would be expected from these trends.}
    \label{fig:nearby}
\end{figure*}

\hiddenref{fig:nearby}

\subsection{Mean polarization}

The error weighted mean of all of the g$^\prime$ observations is $q = -38.6 \pm 0.9$~ppm, $u = -4.6 \pm 0.9$ ppm (or $p = 38.9 \pm 0.9$~ppm). [The mean is higher for MJD 58790 -- 59031 but similar for MJD > 59220]. 
The historic polarization observations described in Section~\ref{sec:intro} support such a small constant component for $\zeta$ Pup. This is surprisingly small for a star at a distance of 332$\pm$11 pc.

\subsubsection{Interstellar polarization}

In general, interstellar polarization increases at a rate of 0.2 to 2~ppm/pc within $\sim$100~pc of the Sun \citep{bailey10, cotton16a}, and at ten times that rate beyond that \citep{behr59}. Indeed, the simple polarization with distance plot of \citet{gontcharov19} shows a median value of $\approx$~3000~ppm, with a minimum of $\approx$~400~ppm for this distance; their Figure 9 suggests the position angle, $\theta$, is likely close to either 90 degrees or 45 degrees (for $\zeta$ Pup this is $93.4 \pm 0.7$ degrees).

To get a more specific understanding of the space around \zPup, in Fig.~\ref{fig:nearby} we have plotted observations of nearby stars from the agglomerated catalogue of \citet{heiles00}. The, largely historical, observations have large uncertainties\footnote{We note that some of these measurements are not significant, and if debiased in the standard way would have $\hat{p}=0$, which would give a false impression of the trend (see \citealp{simmons85}). We have not debiased the data, in part because \citet{heiles00} often assumed a larger error value than obtained. However, this means that the interstellar polarization of the region is likely less than indicated by the raw measurements as plotted in Fig. \ref{fig:nearby}.}, but taken together, the right hand panel indicates a floor in polarization with distance of around 2 ppm/pc, consistent with the value found for Southern hemisphere stars within the Local Hot Bubble by \citet{cotton16a, cotton17b}. However there is a region centred around approximately [116$^\circ$, $-$38$^\circ$] where polarization is seen to increase more steeply -- approaching the 20 ppm/pc found by \citet{behr59} -- beginning from a distance of around 200~pc. The mean polarization of \zPup\ is well below that which would be expected from these trends.

The discrepancy with the expected interstellar polarization implies either multiple misaligned clouds, whose contributions cancel along the line of sight, or a large constant intrinsic polarization component for $\zeta$~Pup -- on the order of 650~ppm -- that cancels the interstellar component. Either scenario is plausible. The former is supported by the bifurcated distribution of polarization position angles for the region indicated by \citet{gontcharov19}.

$\zeta$~Pup seemingly lies on the Sunward side of the centre of the Gum Nebula (centred at [120, $-$43]\footnote{It is hardly probed by stars in the \citet{heiles00} catalogue.}) but within its expanding cloud \citep{woermann01}. It is a similar distance to the Vela OB2 Association \citep{choudhury09}. The Gum Nebula is supposed to be associated with the supernova explosion of $\zeta$~Pup's past companion \citep{choudhury09}. The Gum Nebula spans a few hundred parsecs, and reaches 60~pc past $\zeta$~Pup toward the Sun \citep{woermann01}, and is thus a good candidate for a contrary interstellar component.

\subsubsection{Rotational polarization}

\label{sec:rotpol}

A possible intrinsic mechanism for constant polarization is rapid rotation which results in a net stellar polarization due to the departure of the star from spherical symmetry. Recently this phenomenon has been observed and modelled in a number of stars \citep{cotton17, bailey20a, lewis22, howarth23}. To cancel a presumed interstellar polarization of thousands of ppm, would require a larger effect than has hitherto been observed. 

The methods previously used for modelling the rotational polarization, described in detail in \citet{bailey20a}, require a set of stellar atmosphere models covering the variation in local temperature and gravity from the equator to the pole of the rotating star. As explained in Section~\ref{sec:polmod} the \textsc{atlas9} models we have used previously do not extend to the temperatures and gravities needed for \zPup. In order to allow such modelling to extend to higher temperatures we have obtained the required set of models by interpolating between models in the OSTAR2002 grid as described in Section~\ref{sec:polmod}. However, even this method is not sufficient to reach the gravities required for the equatorial regions of a rapidly rotating \zPup\ model. Such models require some extrapolation beyond the coverage of the OSTAR2002 grid. It should be noted that the grid is limited to values of $\Gamma_{\rm rad}$ < 1 where $\Gamma_{\rm rad}$ is the ratio of radiative to gravitational acceleration \citep[see Section 4 of ][]{lanz2003}. The need to extrapolate the grid therefore means that the Eddington limit is exceeded locally at the equator. 

\begin{figure}
    \centering
    \includegraphics[width=\columnwidth]{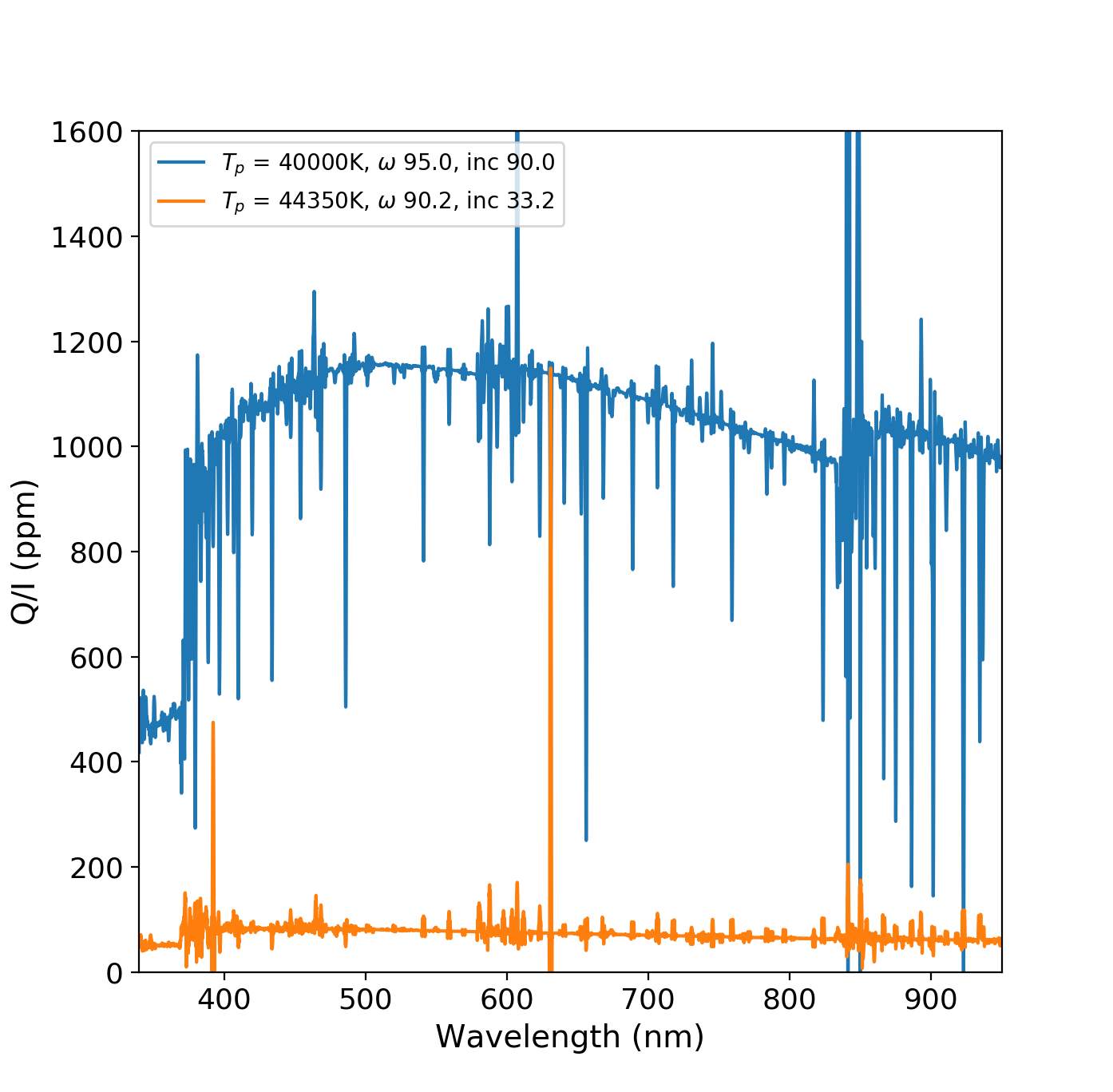}
    \caption{Predicted polarization level due to the rotational distortion of rapidly rotating stars. The lower curve is for the parameters of \zPup\ according to the $M/\msun$ = 25 model of \citet{howarth19}. The upper plot shows the parameters needed to generate a high polarization, in particular a more rapid rotation and a higher inclination, but these parameters are not consistent with what we expect for \zPup.}
    \label{fig:rotating}
\end{figure}

The factors that lead to high polarization are a high rotation rate (specified as $\omega/\omega_c$ where $\omega$ is the equatorial angular velocity of the star, and $\omega_c$ is the critical angular velocity) and a high inclination. High temperature and low gravity also favour high polarization as they increase the relative importance of scattering in the atmosphere. The range of parameters possible for \zPup\ are discussed by \citet{howarth19}. Some of the possible models have slow rotation and are not likely to result in significant rotational polarization. As described in Section~\ref{sec:rotmod}, models that are consistent with $P_{rot} = 1.78$ days are constrained to relatively low inclinations \citep[$i \sim 33$\degr,][]{howarth19} and the $\omega/\omega_c$ depends on the adopted mass. In Fig.~\ref{fig:rotating} we show the rotational polarization predicted for the \citet{howarth19} model with $M/\msun$ = 25 which has $\omega/\omega_c$ = 0.902 and $i$ = 33.2\degr. The polarization is quite low ($\sim$100 ppm) as a consequence of the low inclination. The model with $M/\msun$ = 15 has a larger $\omega/\omega_c$ of 0.985 but the inclination remains low at $i$ = 32.8\degr. This might result in a larger rotational polarization. However, we have not attempted to model this configuration as the low equatorial gravity would require substantial extrapolation beyond the range of the OSTAR2002 grid. As explained above this means that the results of our hydrostatic modelling are unlikely to be meaningful, and such stellar parameters may not be realistic.

The second model plotted in Fig.~\ref{fig:rotating} is an example that shows the conditions needed to get a rotational polarization $\sim$1000 ppm. Both a large $\omega/\omega_c$ and a high inclination are needed and this is not compatible with what we expect for \zPup\ \citep{howarth19}.

\subsubsection{Wind asymmetry}

Net polarization could also result if the wind has an asymmetric shape due to rotation. This possibility was investigated by \citet{harries96} in their analysis of spectropolarimetry of \zPup. These authors had the same problem we have described above, that the interstellar polarization is unknown and so the intrinsic polarization cannot be determined. However they determined a lower limit on the intrinsic polarization of 0.08\% (800 ppm) based on the difference between line and continuum polarization and an upper limit of 0.44\% (4400 ppm) based on their estimate of the maximum likely interstellar polarization. They then used a model of an asymmetric wind and determined that the lower limit corresponded to an equator-to-pole density ratio of 1.3, and the upper limit to a ratio of 3.

Given our different interpretation of the variable polarization in \zPup\ their determined lower limit is no longer valid. However, the value is similar to the $\sim$650 ppm we estimate as a likely interstellar value and so the 1.3 equator-to-pole value is about what might be needed as an asymmetry to cancel such an interstellar polarization. However, it should also be noted that \citet{harries96} used an inclination of 90\degr\ in their modelling, whereas we now think the inclination is $\sim$33\degr\ which will result in smaller polarizations.

If there was a substantial wind asymmetry we would also expect to see an asymmetric distribution in the QU plane for the stochastic variability (lower panel of Fig. \ref{fig:correlation_qu}) since this effectively maps the directions at which clumps are ejected. As discussed in Sections \ref{sec:correlation} and \ref{sec:stochastic} there is no such effect.

\section{Conclusions}

The discovery of previously unobserved polarization variability in a much studied star like \zPup\ is an indication that polarimetry of even the brightest stars is a neglected field. Efficient polarimeters on small telescopes such as Mini-HIPPI on the 23.5-cm telescope used here, or that described by \citet{bailey23}, can make important contributions.

We have made 255 linear polarization observations of \zPup\, including many made at the same time as {\it TESS} observations in early 2021. Spectroscopic observations obtained at the same time show somewhat stronger \Ha\ emission than seen at other times (2000 -- 2016, and 2023). This increase may be related to the increased mass-loss rate from \zPup\ reportedly seen in 2018/2019 Chandra observations \citep{cohen20}.

The polarization is found to show rapid variations on similar timescales to those seen in the photometry. The polarization varies over the photometric 1.78-day period, and also shows more-rapid variability corresponding to the high-frequency stochastic component seen in photometry.

The polarization amplitude ratio (photometric amplitude divided by polarimetric amplitude) is $\sim$12 for the variability as a whole, $\sim$9 for the periodic component and $\sim$19 for the stochastic component. The periodic component shows variation along a preferred position angle $\sim$70\degr\ -- 160\degr. The stochastic variation shows a weak correlation between photometry and polarization and has no preferred direction.

We have tried to fit the 1.78-day variability with a model of a rotating star with bright photospheric spots like that proposed by \citet{rami18}. However, models that fit the light-curve, produce polarization variations with far too low an amplitude and a quite different form of phase curve to those observed.

The presence of polarization variations rules out pulsation in radial ($\ell=0$) or dipole ($\ell=1$) modes as the origin of the 1.78-day periodicity. Non-radial pulsations in $\ell=2$ or higher modes could, in ideal circumstances, produce polarization amplitudes as high as that observed. However, it seems unlikely that such a polarization-favourable mode should be seen in the absence of any other modes. The non-sinusoidal nature and changes of shape of the phase curve also seems inconsistent with pulsation as noted by \citet{rami18}.  

We suggest that a more likely explanation for the polarization variation is scattering from gas in the outflowing wind. This mechanism can more easily produce polarization at the levels observed. Polarization due to a clumpy wind has usually been invoked to explain short timescale polarization variability seen in other hot stars such as Wolf-Rayet stars and OB supergiants (see discussion in Section~\ref{sec:stochastic}). Periodic variations in polarization can be produced by scattering from corotating interaction regions in the wind \citep{ignace15,st-louis18}. The stochastic variability could be explained by a model of randomly ejected clumps like that used for the similar variability observed in WR40 \citep{rami19,ignace23}.

The mean polarization level of \zPup\ is close to zero, which is surprising, considering that we expect significant interstellar polarization at its 332 pc distance. This is presumably the result of fortuitous cancellation of different polarization components with different position angles. The major contributions are most likely interstellar.

\section*{Acknowledgements}

Based in part on data obtained at Siding Spring Observatory. We acknowledge the traditional owners of the land on which the AAT stands, the Gamilaraay people, and pay our respects to elders past and present. DVC thanks the Friends of MIRA for their support.

SLM acknowledges funding support from the Australian Research Council through Discovery Project grant DP180101791 and from the UNSW Scientia Fellowship program. Parts of this work were supported by the Australian Research Council Centre of Excellence for All Sky Astrophysics in 3 Dimensions (ASTRO 3D), through project number CE170100013.

We thank Richard Ignace for comments on the manuscript.

\section*{Data Availability}

The polarization data used for this project are provided in Tables \ref{tab:runs}--\ref{tab:aat_data} of the paper. The {\it TESS} data are in the NASA Mikulski Archive for Space Telescopes (MAST) and were accessed using the Python \textsc{Lightkurve} package.


\bibliographystyle{mnras}
\bibliography{zeta_pup} 




\appendix

\section{Polarization observations and calibration}
\label{sec:polobs}

\hiddenref{tab:runs}
\begin{table*}
\caption{Summary of observing runs and telescope-polarization (TP) calibrations.}
\label{tab:runs}
\tabcolsep 3.5 pt
\begin{tabular}{ll|ccrccc|rcc|rr}
\hline
\multicolumn{2}{c|}{} & \multicolumn{6}{c|}{Telescope and Instrument Set-Up$^a$}   &   \multicolumn{3}{c|}{Observations$^b$}   &   \multicolumn{2}{c}{Calibration}    \\
Run & \multicolumn{1}{c|}{Date Range$^d$} & Instr. &  Tel. & \multicolumn{1}{c}{f/} & Ap. & Mod. & Filter  & $n$ & $\lambda_{\rm eff}$ &  Eff. & \multicolumn{1}{c}{$q_{\text{TP}}$} & \multicolumn{1}{c}{$u_{\rm TP}$} \\
 &  \multicolumn{1}{c|}{(UT)} &  &   &  & ($\arcsec$) &  &  &  & (nm) & ($\%$) & \multicolumn{1}{c}{(ppm)} & \multicolumn{1}{c}{(ppm)} \\
\hline
m2020APR &  2020-04-06 to 2020-06-04  &  M-HIPPI   & 23.5~cm  &   10  & 131.6 &  MT &  $g^{\prime}$   & 57 & 461.3 & 80.7 & 32.4 $\pm$ 3.2 & $-$112.9 $\pm$ 3.2 \\
m2020JUN &  2020-06-16 to 2020-07-01  &  M-HIPPI   & 23.5~cm  &   10  & 131.6 &  MT &  $g^{\prime}$   & 28 & 462.4 & 81.0 & 50.4 $\pm$ 4.7 & $-$132.7 $\pm$ 4.6 \\
m2020JUL &  2020-07-03 to 2020-11-25  &  M-HIPPI   & 23.5~cm  &   10  & 131.6 &  MT &  $g^{\prime}$   & 4 & 461.6 & 80.8 & 45.7 $\pm$ 5.5 & $-$95.0 $\pm$ 5.3 \\
m2021JAN &  2021-01-09 to 2021-03-08  &  M-HIPPI   & 23.5~cm  &   10  & 131.6 &  MT &  $g^{\prime}$   & 129 & 460.2 & 80.4 & 8.9 $\pm$ 2.8 & $-$89.2 $\pm$ 2.8 \\
2021JAN & 2021-01-27 to 2021-01-31  & HIPPI-2  & AAT  & 8*& 11.9 & MBLE1 & $g^\prime$   & 10 & 457.7 & 87.7 & $-$5.5 $\pm$ 1.0 &  $-$5.9 $\pm$ 1.0 \\
&                         &            &      &   &      &        & $r^{\prime}$   & 10 & 601.3 & 63.9 & 25.0 $\pm$ 2.1 & 23.6 $\pm$ 2.3 \\
2021FEB & 2021-02-24 to 2021-02-28  & HIPPI-2  & AAT  & 8*& 11.9 & MBLE1 & 425SP  &  3 &  397.8 & 72.0 & $-$15.6
$\pm$ 1.7 & $-$4.0 $\pm$ 1.9 \\
&                         &            &      &   &      &        & $g^\prime$     & 11 & 458.3 & 87.7 & $-$5.5 $\pm$ 1.0 &  $-$5.9 $\pm$ 1.0 \\
&                         &            &      &   &      &        & $r^{\prime}$   & 3 & 601.5 & 63.9 & 25.0 $\pm$ 2.1 & 23.6 $\pm$ 2.3 \\

\hline
\end{tabular}
\begin{flushleft}
Notes: \\
\textbf{*} Indicates use of a 2$\times$ negative achromatic lens, effectively making the focal ratio f/16.\\ 
\textbf{$^a$} A full description, along with transmission curves for all the components can be found in \citet{bailey20a}. The following parameters were used to calculate modulation efficiency as a function of wavelength; MT: $\lambda_0 =$ 504.5, $Cd =$ 1.726, $e_{\rm max} =$ 0.916; MLBLE1: $\lambda_0 =$ 455.1, $Cd =$ 1.969, $e_{\rm max} =$ 0.926.\\ 
\textbf{$^b$} Mean values are given as representative of the observations made of $\zeta$~Pup, and $n$ is the number of observations.\\
\end{flushleft}
\end{table*}

\begin{table*}
    \centering
    \caption{Polarization observations of $\zeta$~Pup in g$^\prime$ obtained with the Mini-HIPPI polarimeter on the Pindari Observatory 23.5~cm telescope.}
    \begin{tabular}{lrr||lrr||lrr}
    \hline
           MJD & $q$ (ppm) & $u$ (ppm) & MJD & $q$ (ppm) & $u$ (ppm) & MJD & $q$ (ppm) & $u$ (ppm) \\  \hline
58944.535 & $-$240.6$\pm$29.4 & $-$374.3$\pm$28.2 & 59018.351 & $-$262.5$\pm$32.5 & $-$352.9$\pm$31.8 & 59229.520 & $-$251.6$\pm$32.7 & 515.4$\pm$32.6 \\
58972.394 & $-$505.6$\pm$28.0 & $-$ 42.3$\pm$28.1 & 59018.371 & $-$266.1$\pm$31.5 & $-$338.7$\pm$30.9 & 59229.543 & $-$252.5$\pm$34.5 & 440.0$\pm$34.6 \\
58973.388 &  28.3$\pm$30.4 & $-$689.8$\pm$29.3 & 59018.371 & $-$266.1$\pm$31.5 & $-$338.7$\pm$30.9 & 59229.566 & $-$172.9$\pm$38.3 & 357.9$\pm$37.8 \\
58974.368 & $-$244.6$\pm$29.0 & $-$ 66.7$\pm$29.2 & 59018.391 & $-$227.3$\pm$33.0 & $-$416.8$\pm$32.8 & 59229.592 & $-$215.4$\pm$44.0 & 420.0$\pm$44.0 \\
58974.394 & $-$283.5$\pm$29.7 & $-$131.0$\pm$29.5 & 59018.391 & $-$227.3$\pm$33.0 & $-$416.8$\pm$32.8 & 59229.643 & $-$ 70.5$\pm$41.5 & 514.0$\pm$41.9 \\
58974.415 & $-$213.8$\pm$27.9 & $-$164.2$\pm$28.0 & 59019.355 & $-$ 54.6$\pm$31.4 & $-$267.3$\pm$31.4 & 59229.665 & $-$ 86.1$\pm$40.0 & 488.7$\pm$40.1 \\
58974.437 & $-$200.0$\pm$30.2 & $-$249.7$\pm$30.3 & 59019.374 & $-$ 72.0$\pm$31.2 & $-$155.0$\pm$32.0 & 59229.687 & $-$119.9$\pm$37.0 & 475.0$\pm$36.6 \\
58974.490 & $-$143.2$\pm$32.2 & $-$245.6$\pm$32.2 & 59019.392 & $-$177.5$\pm$31.7 & $-$ 24.3$\pm$33.3 & 59229.713 &  25.1$\pm$32.2 & 445.9$\pm$37.9 \\
58974.512 & $-$328.1$\pm$34.5 & $-$118.4$\pm$34.6 & 59022.361 & $-$158.3$\pm$31.0 &  35.3$\pm$29.6 & 59229.736 & 140.2$\pm$34.5 & 269.0$\pm$35.0 \\
58975.371 & $-$160.8$\pm$45.0 &  60.1$\pm$45.1 & 59022.384 & $-$159.2$\pm$31.8 & $-$ 38.4$\pm$31.8 & 59230.463 & 137.1$\pm$45.0 & $-$330.4$\pm$32.6 \\
58980.393 & 418.8$\pm$28.5 & $-$281.5$\pm$27.4 & 59023.335 & 199.7$\pm$28.5 & 215.5$\pm$37.8 & 59230.485 & 109.1$\pm$28.5 & $-$307.2$\pm$29.5 \\
58980.448 & 588.6$\pm$29.4 & $-$613.3$\pm$29.9 & 59023.354 &  42.3$\pm$29.4 & 122.9$\pm$40.1 & 59230.507 & 266.6$\pm$29.4 & $-$217.9$\pm$31.6 \\
58982.477 & 185.4$\pm$31.7 & 107.7$\pm$31.9 & 59023.374 &  24.0$\pm$31.7 &  68.9$\pm$37.7 & 59230.530 & 150.2$\pm$31.7 &  52.6$\pm$31.4 \\
58985.362 & $-$732.4$\pm$28.2 & 190.3$\pm$27.4 & 59023.393 & $-$ 46.5$\pm$37.6 & $-$ 50.1$\pm$35.2 & 59230.555 & 150.1$\pm$28.2 & 156.2$\pm$30.1 \\
58985.384 & $-$588.8$\pm$29.3 & 184.3$\pm$27.3 & 59025.337 & 512.8$\pm$29.3 &  85.8$\pm$30.1 & 59230.580 & $-$ 89.4$\pm$30.5 & 231.8$\pm$30.4 \\
58985.404 & $-$404.0$\pm$30.0 & 210.0$\pm$28.8 & 59025.356 & 354.1$\pm$30.0 & 197.2$\pm$30.9 & 59230.603 & $-$163.5$\pm$31.8 & 179.9$\pm$31.8 \\
58985.425 & $-$286.5$\pm$30.5 & 246.9$\pm$30.0 & 59025.376 & 157.2$\pm$30.5 & 186.2$\pm$30.3 & 59230.627 & $-$ 35.2$\pm$30.2 & 246.8$\pm$30.1 \\
58985.449 & $-$ 55.3$\pm$30.4 & 196.3$\pm$31.5 & 59026.338 & $-$ 86.8$\pm$31.7 & $-$346.1$\pm$35.9 & 59230.650 & $-$ 64.6$\pm$31.5 & 110.3$\pm$31.5 \\
58985.470 & $-$  8.2$\pm$32.4 &  55.6$\pm$32.7 & 59026.358 & $-$ 94.6$\pm$32.4 & $-$278.1$\pm$33.2 & 59230.672 &  56.6$\pm$32.4 & 119.0$\pm$32.1 \\
58985.495 &  57.6$\pm$37.8 & $-$249.6$\pm$42.5 & 59026.376 & $-$174.0$\pm$32.3 & $-$175.8$\pm$31.5 & 59230.694 &  44.9$\pm$37.8 &  79.0$\pm$32.6 \\
58986.402 & $-$ 67.0$\pm$31.4 & 140.1$\pm$30.4 & 59029.352 & 350.1$\pm$31.4 & $-$668.5$\pm$32.9 & 59230.717 & 107.0$\pm$31.4 &  73.6$\pm$32.9 \\
58986.437 & $-$299.8$\pm$30.4 &  36.3$\pm$30.4 & 59029.372 & 242.2$\pm$30.4 & $-$630.4$\pm$33.4 & 59232.464 &  74.7$\pm$30.4 & 115.6$\pm$33.6 \\
58986.491 & $-$350.6$\pm$34.1 & $-$175.8$\pm$32.0 & 59030.339 & 203.9$\pm$34.1 & $-$247.2$\pm$29.4 & 59232.489 & $-$ 53.1$\pm$32.1 &  74.2$\pm$32.0 \\
58988.345 & $-$133.5$\pm$30.0 & $-$506.3$\pm$29.4 & 59030.359 & 235.8$\pm$30.0 & $-$256.9$\pm$31.2 & 59232.512 &  14.6$\pm$30.0 &  71.4$\pm$33.3 \\
58988.367 & $-$273.5$\pm$28.5 & $-$333.2$\pm$29.6 & 59030.379 & 297.7$\pm$28.5 & $-$158.4$\pm$33.0 & 59232.534 &  99.1$\pm$28.5 & 128.7$\pm$32.7 \\
58989.346 & 100.5$\pm$28.7 & $-$636.9$\pm$29.4 & 59033.340 & 137.4$\pm$28.7 & $-$337.1$\pm$33.2 & 59232.559 & 246.9$\pm$28.7 & 114.2$\pm$32.4 \\
58989.373 & 301.3$\pm$32.5 & $-$591.5$\pm$29.2 & 59033.360 & $-$179.1$\pm$36.4 & $-$ 97.7$\pm$36.8 & 59232.581 & 288.1$\pm$32.5 & 152.0$\pm$32.7 \\
58989.401 & 449.8$\pm$28.8 & $-$600.0$\pm$29.1 & 59178.630 & $-$ 92.8$\pm$34.4 &  66.6$\pm$34.9 & 59232.611 & 164.9$\pm$28.8 & $-$  3.1$\pm$32.9 \\
58989.422 & 461.3$\pm$28.7 & $-$493.6$\pm$30.0 & 59178.651 & $-$ 96.0$\pm$34.7 &  28.1$\pm$34.6 & 59232.633 & 132.6$\pm$28.7 &  55.3$\pm$34.4 \\
58997.357 & $-$482.3$\pm$29.9 &  41.2$\pm$30.4 & 59223.523 & $-$268.4$\pm$30.9 & $-$ 16.7$\pm$30.9 & 59232.655 & 217.6$\pm$29.9 &   1.2$\pm$34.4 \\
58997.378 & $-$440.6$\pm$29.3 & $-$ 15.8$\pm$30.3 & 59224.484 & $-$365.7$\pm$32.0 &  87.7$\pm$32.1 & 59234.461 & $-$271.5$\pm$33.4 & $-$240.3$\pm$33.1 \\
58997.400 & $-$457.4$\pm$30.9 & $-$ 64.9$\pm$33.6 & 59224.507 & $-$328.7$\pm$32.6 &  94.9$\pm$32.5 & 59234.483 & $-$238.7$\pm$33.4 & $-$284.0$\pm$33.4 \\
58997.424 & $-$290.7$\pm$32.2 & $-$155.6$\pm$33.0 & 59224.530 & $-$132.0$\pm$31.8 &  89.3$\pm$31.8 & 59234.508 & $-$272.9$\pm$32.0 & $-$354.2$\pm$32.0 \\
58997.454 & $-$191.8$\pm$32.6 & $-$140.1$\pm$34.9 & 59224.552 &  20.4$\pm$32.6 & $-$ 77.6$\pm$32.4 & 59234.533 & $-$288.5$\pm$32.4 & $-$327.4$\pm$32.2 \\
58998.367 &  56.4$\pm$32.6 & $-$386.8$\pm$30.2 & 59224.575 & $-$ 16.7$\pm$31.8 & $-$297.4$\pm$31.8 & 59234.562 & $-$450.2$\pm$32.9 & $-$189.3$\pm$32.9 \\
58998.425 &  35.4$\pm$32.7 & $-$249.0$\pm$33.7 & 59224.597 & $-$ 51.8$\pm$32.1 & $-$337.2$\pm$32.2 & 59234.597 & $-$649.4$\pm$32.7 & 106.2$\pm$32.8 \\
58998.452 &  89.8$\pm$34.2 & $-$103.2$\pm$35.5 & 59225.494 & 159.7$\pm$34.2 & $-$287.0$\pm$31.1 & 59234.624 & $-$690.8$\pm$33.3 & 172.9$\pm$33.3 \\
58999.340 & $-$191.2$\pm$42.9 &  31.0$\pm$43.2 & 59225.518 & 276.4$\pm$42.9 & $-$251.5$\pm$32.1 & 59234.648 & $-$800.1$\pm$33.1 & 101.8$\pm$33.1 \\
58999.384 & $-$168.4$\pm$30.8 & 297.9$\pm$30.7 & 59225.543 & 343.3$\pm$30.8 & $-$ 86.1$\pm$31.8 & 59234.671 & $-$797.6$\pm$33.6 &  45.6$\pm$33.5 \\
58999.427 & $-$342.6$\pm$32.2 & 285.8$\pm$32.1 & 59225.566 & 222.4$\pm$32.2 &  59.3$\pm$30.8 & 59235.464 & 348.8$\pm$32.2 & $-$348.7$\pm$32.9 \\
58999.447 & $-$350.8$\pm$32.8 & 192.2$\pm$32.8 & 59225.592 & 150.0$\pm$32.8 & 182.9$\pm$32.6 & 59235.488 & 535.3$\pm$32.8 & $-$347.0$\pm$33.7 \\
59000.384 &  83.2$\pm$32.4 & $-$678.5$\pm$32.4 & 59225.615 & 124.0$\pm$32.4 & 125.3$\pm$32.3 & 59235.512 & 641.2$\pm$32.4 & $-$315.8$\pm$33.7 \\
59000.405 &  36.8$\pm$32.2 & $-$437.6$\pm$32.2 & 59225.639 & 123.2$\pm$32.2 & 109.8$\pm$31.5 & 59235.536 & 578.9$\pm$32.2 & $-$453.5$\pm$32.4 \\
59000.440 &  60.9$\pm$44.8 & $-$285.7$\pm$43.3 & 59225.661 & 135.3$\pm$44.8 & 145.5$\pm$32.1 & 59235.561 & 635.2$\pm$44.8 & $-$414.0$\pm$32.8 \\
59001.417 & $-$783.6$\pm$31.7 & 152.8$\pm$32.1 & 59225.686 &  62.0$\pm$31.7 & 125.4$\pm$31.9 & 59237.462 & 336.4$\pm$31.7 &  53.1$\pm$33.3 \\
59001.452 & $-$731.8$\pm$34.0 & $-$127.8$\pm$34.3 & 59225.711 & 202.3$\pm$34.0 & 188.1$\pm$33.9 & 59237.486 & 420.2$\pm$34.0 & $-$ 22.7$\pm$32.3 \\
59002.336 & 367.9$\pm$28.5 & $-$279.3$\pm$28.2 & 59225.736 & 187.8$\pm$28.5 & 106.4$\pm$34.7 & 59237.511 & 500.1$\pm$28.5 &  80.2$\pm$32.9 \\
59002.358 & 436.0$\pm$29.6 & $-$284.1$\pm$29.7 & 59226.470 & 576.6$\pm$29.6 & $-$493.4$\pm$32.7 & 59237.534 & 315.4$\pm$29.6 &  36.4$\pm$32.7 \\
59002.380 & 323.3$\pm$29.4 & $-$191.3$\pm$29.3 & 59226.494 & 630.9$\pm$29.4 & $-$287.7$\pm$32.6 & 59237.559 & 239.4$\pm$29.4 & $-$ 15.8$\pm$33.7 \\
59002.428 & 141.1$\pm$32.0 &  44.4$\pm$32.1 & 59226.519 & 650.7$\pm$32.0 & $-$329.1$\pm$33.6 & 59251.456 & 128.4$\pm$32.0 & $-$282.5$\pm$33.2 \\
59002.450 & 204.0$\pm$33.5 & $-$ 82.9$\pm$33.7 & 59226.542 & 583.2$\pm$33.5 & $-$225.3$\pm$33.5 & 59251.481 & 158.7$\pm$33.5 & $-$279.4$\pm$33.8 \\
59003.337 & $-$282.5$\pm$30.1 & $-$ 10.1$\pm$30.1 & 59226.565 & 431.8$\pm$30.1 & $-$234.1$\pm$33.3 & 59251.504 & 115.3$\pm$30.1 & $-$233.5$\pm$33.8 \\
59003.359 & $-$146.6$\pm$32.1 &  19.7$\pm$32.2 & 59226.588 & 381.3$\pm$32.1 & $-$192.5$\pm$32.8 & 59251.528 & 384.7$\pm$32.1 & $-$269.7$\pm$34.7 \\
59004.339 & $-$146.9$\pm$30.6 & 400.0$\pm$30.4 & 59227.475 & $-$ 91.9$\pm$35.1 & 528.5$\pm$35.4 & 59251.553 & 478.9$\pm$30.6 & $-$320.2$\pm$36.6 \\
59004.361 &  91.3$\pm$31.2 & 309.4$\pm$30.9 & 59227.501 & $-$153.4$\pm$36.5 & 421.2$\pm$36.0 & 59251.577 & 561.8$\pm$31.2 & $-$412.3$\pm$37.6 \\
59004.383 & 281.0$\pm$32.1 & 282.4$\pm$31.8 & 59227.539 & $-$267.9$\pm$32.8 & 584.2$\pm$32.7 & 59251.599 & 518.1$\pm$32.1 & $-$408.9$\pm$35.8 \\
59004.406 & 456.3$\pm$33.7 & 390.1$\pm$33.3 & 59227.562 & $-$231.9$\pm$34.7 & 532.5$\pm$34.6 & 59251.622 & 433.4$\pm$33.7 & $-$505.2$\pm$41.1 \\
59016.366 & $-$347.7$\pm$34.0 & $-$205.5$\pm$30.5 & 59227.601 & $-$ 97.7$\pm$35.1 & 566.4$\pm$35.1 & 59251.663 & 325.4$\pm$34.0 & $-$415.4$\pm$37.3 \\
59016.390 & $-$484.8$\pm$34.5 & $-$422.6$\pm$32.5 & 59229.467 & $-$235.2$\pm$30.9 & 697.5$\pm$31.2 & 59256.426 & 455.6$\pm$34.5 & 311.1$\pm$35.7 \\
59018.351 & $-$262.5$\pm$32.5 & $-$352.9$\pm$31.8 & 59229.494 & $-$279.0$\pm$33.5 & 768.0$\pm$33.4 & 59256.450 & 436.7$\pm$32.5 & 287.6$\pm$36.0 \\
\hline
    \end{tabular}
    \label{tab:mhippi_data}
\end{table*}

\begin{table*}

    \centering
    \contcaption{}
    \begin{tabular}{lrr||lrr||lrr}
    \\
    \hline
           MJD & $q$ (ppm) & $u$ (ppm) & MJD & $q$ (ppm) & $u$ (ppm) & MJD & $q$ (ppm) & $u$ (ppm) \\  \hline
59256.472 & 346.7$\pm$35.8 & 227.8$\pm$35.9 & 59264.595 & 381.7$\pm$35.8 &  94.1$\pm$37.4 & 59277.502 & $-$267.1$\pm$33.0 & $-$112.9$\pm$33.0 \\
59259.424 & 117.4$\pm$34.9 & 212.7$\pm$35.0 & 59264.631 & 396.0$\pm$34.9 & 262.1$\pm$37.6 & 59277.524 & $-$211.5$\pm$33.5 & $-$  7.4$\pm$33.5 \\
59259.448 & $-$110.0$\pm$33.0 &  22.2$\pm$33.2 & 59265.431 &  34.1$\pm$33.0 & 346.8$\pm$36.5 & 59277.546 & $-$313.8$\pm$35.4 & 174.3$\pm$35.2 \\
59259.477 & $-$284.0$\pm$35.8 & $-$ 25.9$\pm$35.8 & 59265.461 & $-$199.0$\pm$36.2 & 122.4$\pm$36.3 & 59277.567 & $-$295.2$\pm$34.8 & 413.2$\pm$34.7 \\
59259.500 & $-$334.3$\pm$33.9 & $-$ 93.3$\pm$33.9 & 59265.484 & $-$ 39.6$\pm$34.9 & $-$ 11.8$\pm$34.9 & 59277.588 & $-$358.0$\pm$34.5 & 453.2$\pm$34.7 \\
59259.524 & $-$409.6$\pm$32.5 & $-$ 54.0$\pm$32.4 & 59265.509 & 218.0$\pm$32.5 & $-$ 68.3$\pm$37.8 & 59280.405 & 157.8$\pm$36.1 & $-$346.5$\pm$35.9 \\
59259.548 & $-$423.7$\pm$30.7 & $-$163.5$\pm$31.0 & 59266.425 & 297.2$\pm$30.7 &  41.6$\pm$36.1 & 59280.428 & 300.9$\pm$37.7 & $-$437.1$\pm$37.9 \\
59264.431 & $-$173.6$\pm$34.1 &  88.9$\pm$34.0 & 59266.450 & 290.9$\pm$34.1 & $-$ 87.0$\pm$34.2 & 59280.451 & 464.1$\pm$37.4 & $-$495.2$\pm$37.0 \\
59264.454 & $-$242.0$\pm$33.6 &  62.8$\pm$33.5 & 59266.501 & 106.4$\pm$33.6 & $-$180.3$\pm$35.3 & 59281.406 & 310.3$\pm$39.1 & 110.4$\pm$38.5 \\
59264.477 & $-$151.1$\pm$35.0 &  57.0$\pm$34.9 & 59277.408 & $-$ 68.8$\pm$34.1 & 290.6$\pm$34.0 & 59281.431 & 238.5$\pm$35.2 & 275.9$\pm$35.1 \\
59264.504 & $-$ 57.6$\pm$34.7 & $-$  3.7$\pm$34.8 & 59277.432 & $-$178.4$\pm$32.8 & 197.6$\pm$32.8 & 59281.454 & 151.6$\pm$35.4 & 234.2$\pm$35.4 \\
59264.548 & 139.5$\pm$35.1 & $-$173.4$\pm$34.9 & 59277.454 & $-$218.4$\pm$33.6 &  33.7$\pm$33.7 & 59281.479 & 250.9$\pm$37.2 & 172.7$\pm$36.9 \\
59264.571 & 321.9$\pm$35.5 & $-$101.3$\pm$35.5 & 59277.478 & $-$203.2$\pm$33.7 & $-$  4.3$\pm$33.5& & &  \\
\hline
    \end{tabular}
\end{table*}

\begin{table}
    \centering
    \caption{Multi-filter polarization observations of $\zeta$~Pup with HIPPI-2 on the AAT.}
    \begin{tabular}{lrrr}
    \hline
     MJD & Fil & $q$ (ppm) & $u$ (ppm) \\ \hline
59241.460 & g$^\prime$ & 60.0$\pm$\phantom{0}6.7 & $-$ 48.8$\pm$\phantom{0}5.7 \\
59241.466 & r$^\prime$ & 71.5$\pm$17.4 & $-$ 12.9$\pm$19.4 \\
59241.548 & g$^\prime$ & $-$ 71.5$\pm$\phantom{0}4.1 & 137.6$\pm$\phantom{0}3.9 \\
59241.553 & r$^\prime$ & $-$ 41.3$\pm$14.4 & 203.1$\pm$18.2 \\
59241.626 & g$^\prime$ & $-$496.7$\pm$\phantom{0}4.5 & 423.9$\pm$\phantom{0}4.1 \\
59241.631 & r$^\prime$ & $-$539.2$\pm$13.2 & 452.0$\pm$13.2 \\
59242.518 & g$^\prime$ & 429.6$\pm$\phantom{0}4.3 & $-$181.5$\pm$\phantom{0}4.5 \\
59242.529 & r$^\prime$ & 385.6$\pm$12.1 & $-$103.0$\pm$13.1 \\
59242.622 & g$^\prime$ & 477.1$\pm$\phantom{0}4.3 &  51.6$\pm$\phantom{0}3.8 \\
59242.632 & r$^\prime$ & 501.4$\pm$22.4 & 145.3$\pm$18.5 \\
59244.557 & g$^\prime$ & 645.3$\pm$\phantom{0}7.3 & $-$449.6$\pm$\phantom{0}6.8 \\
59244.563 & r$^\prime$ & 520.6$\pm$21.7 & $-$396.3$\pm$22.7 \\
59244.716 & g$^\prime$ & 655.3$\pm$\phantom{0}4.6 & $-$453.2$\pm$\phantom{0}4.1 \\
59244.722 & r$^\prime$ & 514.5$\pm$14.3 & $-$393.7$\pm$12.6 \\
59245.458 & g$^\prime$ & $-$661.0$\pm$\phantom{0}3.7 &  27.5$\pm$\phantom{0}3.9 \\
59245.463 & r$^\prime$ & $-$583.3$\pm$13.8 & 108.0$\pm$14.3 \\
59245.577 & g$^\prime$ & $-$227.1$\pm$\phantom{0}5.9 & 208.1$\pm$\phantom{0}6.0 \\
59245.584 & r$^\prime$ & $-$209.4$\pm$13.1 & 238.8$\pm$14.8 \\
59245.702 & g$^\prime$ & $-$304.8$\pm$\phantom{0}8.6 & 398.9$\pm$\phantom{0}9.6 \\
59245.709 & r$^\prime$ & $-$243.2$\pm$17.6 & 440.6$\pm$19.8 \\
59269.571 & g$^\prime$ & 201.2$\pm$\phantom{0}3.9 & $-$ 80.3$\pm$\phantom{0}4.1 \\
59269.668 & g$^\prime$ & 506.6$\pm$\phantom{0}6.0 & $-$246.5$\pm$\phantom{0}5.1 \\
59270.401 & g$^\prime$ & $-$385.2$\pm$\phantom{0}6.7 & $-$151.4$\pm$\phantom{0}9.5 \\
59270.461 & g$^\prime$ & $-$445.3$\pm$\phantom{0}3.6 &   8.8$\pm$\phantom{0}3.7 \\
59270.466 & 425SP & $-$531.6$\pm$13.0 &  11.2$\pm$13.1 \\
59270.475 & r$^\prime$ & $-$371.0$\pm$13.2 &  42.0$\pm$16.1 \\
59270.555 & g$^\prime$ & $-$202.2$\pm$\phantom{0}4.0 & 202.2$\pm$\phantom{0}3.7 \\
59270.624 & g$^\prime$ & $-$108.2$\pm$\phantom{0}3.9 &  85.5$\pm$\phantom{0}3.5 \\
59270.679 & g$^\prime$ & $-$ 89.8$\pm$\phantom{0}3.8 & 165.3$\pm$\phantom{0}3.9 \\
59272.596 & g$^\prime$ & 217.6$\pm$\phantom{0}4.1 & $-$285.0$\pm$\phantom{0}4.1 \\
59272.654 & g$^\prime$ & 212.7$\pm$\phantom{0}4.2 & $-$374.4$\pm$\phantom{0}4.0 \\
59272.660 & 425SP & 231.9$\pm$12.8 & $-$439.3$\pm$13.0 \\
59272.667 & r$^\prime$ & 140.7$\pm$10.5 & $-$259.5$\pm$10.5 \\
59273.670 & g$^\prime$ & $-$324.8$\pm$\phantom{0}4.1 & 100.6$\pm$\phantom{0}4.2 \\
59273.699 & g$^\prime$ & $-$350.1$\pm$\phantom{0}4.3 & 341.7$\pm$\phantom{0}4.3 \\
59273.705 & 425SP & $-$357.4$\pm$13.2 & 489.6$\pm$13.2 \\
59273.711 & r$^\prime$ & $-$350.8$\pm$12.8 & 375.3$\pm$12.5 \\
\hline
    \end{tabular}
    \label{tab:aat_data}
\end{table}

The Mini-HIPPI instrument used here includes a number of modifications from that described by \citet{bailey17}. These include the addition of a six-position filter wheel and a new compact electronics unit based on the same design as used for HIPPI-2 \citep{bailey20a}.  The data were reduced using the methods described by \citet{bailey20a}. Mini-HIPPI  was previously used, on a different telescope, to discover the phase-dependent polarization variability of the binary systems Spica \citep{bailey19a} and $\mu^1$ Sco \citep{cotton20}.

Each observation requires measurements of the star at four instrument position angles (0$\degr$, 45$\degr$, 90$\degr$ and 135$\degr$), together with sky measurements at the same four angles. For the Mini-HIPPI observations the total exposure time on star was 800 seconds, but due to overheads in the process of moving to sky and recentering the star, which was inefficient on the small telescope, a typical observation lasted from 30 to 35 minutes. For the HIPPI-2 observations exposure times were 160 seconds in the $g^\prime$ filter, 320 seconds in $r^\prime$, and 240 seconds in 425SP.

Table~\ref{tab:runs} summarizes the polarization observing runs. The telescope polarization (TP) was calibrated using observations of low-polarization standard stars as described in \citet{bailey20a}. The low-polarization standards used for the Mini-HIPPI observations were Sirius and $\alpha$~Cen. The measured telescope-polarization values for each run  are listed in 
Table~\ref{tab:runs} (as $q_{\rm TP}$, $u_{\rm TP}$).

The polarization position-angle for HIPPI-2 was calibrated using observations of the polarized standard stars listed in Table~4 of \citet{bailey20a}. For Mini-HIPPI observations, we used two bright standards from that list (HD~84810 and HD~187929) and an additional standard, HD~161471 ($\iota^1$~Sco, 
$V = 3.0$) for which we adopt the parameters 
$p_{\rm max} = 2.28\%$, $\lambda_{\rm max} = 0.56~\mu$m, $\theta = 2.8^\circ$ \citep{serkowski75}.

The polarization data for the Mini-HIPPI observations are listed in Table~\ref{tab:mhippi_data}. Polarization values are given as normalized Stokes parameters $q = Q/I$ and $u = U/I$, measured in parts per million (ppm). Observing times are given as the Modified Julian Date (MJD = JD~$-$~2400000.5) for the mid-point of the observation, corrected to the Solar-system barycentre. 
Table~\ref{tab:aat_data} lists, in a similar format, the data obtained with HIPPI-2 on the AAT. The quoted uncertainties include the instrumental positioning error (14.0 ppm for the Mini-HIPPI observations), as discussed in \citet{bailey20a}.

In an upcoming work (Cotton et al., in prep.) we redetermine the parameters that characterise the efficiency of the HIPPI-2 and Mini-HIPPI modulators based on a decade of on-sky measurements, rather than a mix of laboratory and on-sky determinations as presented in \citet{bailey17, bailey20a}. The main impact of this is to shift the maximum efficiency $e_{\rm max}$ of the different units by several percent. The difference is significant with respect to the reported errors for objects with large polarizations, as presented here. Consequently, here we adopt interim efficiency parameters, based on the work in progress; these are given in a footnote to Table~\ref{tab:runs}. For the MT modulator used with Mini-HIPPI, the reduced efficiency of the Glan-Taylor prism compared to the Wollaston prism, used with HIPPI-2, is wrapped into these figures.



\bsp	
\label{lastpage}
\end{document}